\documentclass[preprint,12pt]{elsarticle}
\usepackage{epsfig,rotating,setspace,latexsym,amsmath,epsf,amssymb,bm,amsbsy}
\usepackage{graphicx,color,subfigure,here}
\usepackage[normalem]{ulem} 
\usepackage{algorithm}
\usepackage[noend]{algpseudocode}
\usepackage{epstopdf}
\usepackage{scrextend}
\usepackage{ifpdf}
\usepackage{dsfont}
\usepackage[english]{babel}
\usepackage{multirow}
\newcommand{\bengi}[1]{{\leavevmode\color{black}#1}} 

\journal{Computer Communications}

\begin{document}

\begin{frontmatter}

\title{ECPR: Environment-and Context-aware Combined Power and Rate Distributed Congestion Control for Vehicular Communications}

\author[wpi]{Bengi Aygun\corref{mycorrespondingauthor}}\ead{baygun@wpi.edu}  \cortext[mycorrespondingauthor]{Corresponding author} 
\author[huawei]{Mate Boban\corref{NEC}}\ead{mate.boban@huawei.com}\cortext[NEC]{M. Boban carried out this work while at NEC Laboratories Europe.} 
\author[wpi]{Alexander M. Wyglinski} \ead{alexw@wpi.edu} 
\address[wpi]{Wireless Innovation Laboratory, Department of Electrical and Computer Engineering\\
Worcester Polytechnic Institute, Worcester, MA 01609-2280, USA}
\address[huawei]{Huawei European Research Center, Munich, Germany}

\begin{abstract}
Safety and efficiency applications in vehicular networks rely on the exchange of periodic messages between vehicles. These messages contain position, speed, heading, and other vital information that makes the vehicles aware of their surroundings. The drawback of exchanging periodic cooperative messages is that they generate significant channel load.  Decentralized Congestion Control (DCC) algorithms have been proposed to minimize the channel load.  However, while the rationale for periodic message exchange is \emph{to improve awareness}, existing DCC algorithms do not use awareness as a metric for deciding when, at what power, and at what rate the periodic messages need to be sent in order to make sure all vehicles are informed. We propose an environment- and context-aware DCC algorithm combines power and rate control in order to improve cooperative awareness by adapting to both specific propagation environments (\textit{e.g.}, urban intersections, open highways, suburban roads) as well as application requirements (\textit{e.g.}, different target cooperative awareness range). Studying various operational conditions (\textit{e.g.}, speed, direction, and application requirement), ECPR adjusts the transmit power of the messages in order to reach the desired awareness ratio at the target distance while at the same time controlling the channel load using an adaptive rate control algorithm. By performing extensive simulations, including realistic propagation as well as environment modeling and realistic vehicle operational environments (varying demand on both awareness range and rate), we show that ECPR can increase awareness by $20\%$ while keeping the channel load and interference at almost the same level. When permitted by the awareness requirements, ECPR can improve the average message rate by $18\%$ compared to algorithms that perform rate adaptation only.
\end{abstract}
\end{frontmatter}

\section{Introduction} \label{sec:Intro}
The U.S. Department of Transportation announced that connected road vehicles will be mandated by 2017~\cite{nhtsa}.  As such, wireless communication technologies have been studied in order to enable reliable connected vehicles across any of operating conditions. One promising solution is vehicular ad hoc networks (VANETs), which has been actively studied over past several decades\bengi{~\cite{campolo15,campolo15_2, chen15}}. The key building block for enabling many safety applications in VANETs is cooperative awareness. The main premise for cooperative awareness is that by knowing their operating environment, vehicles and their drivers are going to be better equipped for decision-making in hazardous situations (\textit{e.g.}, emergency braking) and more adept at finding better routes to their destination (\textit{e.g.}, avoiding congested roads). To enable cooperative awareness, vehicles use periodic message exchanges (also referred to as ``beaconing'') in order to exchange position, speed, heading, and other vital information that makes the vehicles aware of their surroundings. Such cooperative awareness is used to enable safety applications, such as intersection collision warning, accident warning, and emergency braking~\cite{etsi09}. Since they are sent periodically by all vehicles, beacons are envisioned to occupy a large proportion of the channel time~\cite{etsi14TR101612}. Decentralized Congestion Control (DCC) algorithms can be used to control the number of beacons and other messages transmitted across the channel. Typically, DCC approaches in VANETs are classified as: 1) rate control; 2) power control; and 3) combined rate and power control. Rate control algorithms adapt the message rate, \textit{i.e.} number of packets per unit time that a vehicle can transmit, where the rate is often adapted based on the channel load information. Power adaptation algorithms use transmit power control to limit the range over which a message is broadcast, thus effectively controlling the channel load. Combined algorithms employ the previous two types of control by applying both rate control to reduce the number of messages and power control to limit their range.

In recent years, there have been a number of works on DCC approaches proposed for VANETs. Since the standardization of DCC is vital for interoperability and performance of vehicle-to-X (V2X) communications, there continues to be ongoing research on DCC in various standardization bodies and special interests consortia (\textit{e.g.}, within European Telecommunications Standards Institute (ETSI) and as part of the Car-to-Car Communications Consortium) aimed at performance evaluation and providing a unified cross-layer DCC framework~\bengi{\cite{etsi14TR101612, puthalIccve13, zemouriItsc14, shagdar15, sepulcre11}}. One example of a metric that is often used is the channel busy ratio (CBR), defined as the proportion of channel time that is deemed occupied by an ongoing transmission. Bansal \textit{et al.} devised an algorithm called the LInear MEssage Rate Integrated Control (LIMERIC)~\cite{bansal}, a rate control algorithm that adapts the message rate by using CBR measurements in a linear manner (\textit{e.g.}, proportional to the change of CBR). The authors prove that the convergence of LIMERIC yields fair and efficient channel utilization. Tielert \textit{et al.}~\cite{tielert} proposed an algorithm called PULSAR (Periodically Updated Load Sensitive Adaptive Rate control), which uses piggybacked two-hop CBR information and additive increase multiplicative decrease method (AIMD) in order to achieve better channel utilization and max-min fairness. The approaches described above used linear rate adaptation. A simpler approach to rate control is to increase/decrease the rate based on, for example, the CBR being above or below a preset threshold. This approach is frequently referred to as binary rate control. One example of a binary rate control algorithm is Context-Aware Rate Selection (CARS) by Shankar \textit{et al.}~\cite{shankar08}. Egea-Lopez and Pavon-Marino~\cite{egea2014distributed} reformulated the congestion control problem as a network utility maximization problem and design fair adaptive beaconing rate for intervehicular communications (FABRIC), a proportionally fair binary rate control algorithm. \bengi{The required message rate may change depending on the situation. To deal with these differences, Joerer \textit{et al.}~\cite{joerer15} perform rate adaptation by considering the context.}

Power adaptation algorithms use transmit power control to limit the range over which a message is broadcasted, thus effectively controlling the channel load. Torrent-Moreno \textit{et al.}~\cite{moreno09} designed a power control algorithm aimed at ensuring bandwidth allocation for high-priority event-based messages (\textit{e.g.}, for safety applications), whereas Mittag \textit{et al.}~\cite{mittag2008} elaborated on the same algorithm by introducing segment-based power control with the goal of reducing overhead. By testing the solution on homogeneous vehicular traffic densities and imperfect channel information, the authors demonstrated the effectiveness of their algorithm. Caizzone \textit{et al.}~\cite{caizzone} proposed an algorithm that adapts transmit power depending on the number of neighbors, where the transmit power is increased in case the number of neighbors is under the threshold or \textit{vice versa}. Regarding combined power and rate adaptation algorithms, Le \textit{et al.}~\cite{le2011performance} evaluated rate-only, power-only, and combined rate and power control algorithms. By performing extensive simulations, the authors identified which of the algorithms is preferable for a specific scenario and application requirement. \bengi{Kloiber \textit{et al.}~\cite{kloiber16} introduced a random transmission power assignment  in order to make correlated packet collisions more uncorrelated in space. Authors in~\cite{gomez16, math15, marzouk15} define DCC problem as a state machine to perform transmission power control.} Khorakhun \textit{et al.}~\cite{khorakhun2008} combined the binary rate adaptation with transmit power control, where the increase/decrease of transmit power is defined with a parameter chosen based on CBR. Tielert \textit{et al.}~\cite{tielert13} adapted the transmit power and rate with respect to the target transmission distance and channel conditions by using Pareto optimal parameter combinations. The authors point out that there is a need for further study involving variable channel conditions, including dynamic transitions between line-of-sight (LOS) and non-LOS conditions, which was experimentally shown to have a profound impact on communication performance, and with significant real-world effect on congestion control algorithms~\cite{schmidt11}. 

Since congestion control is inherently a cross-layer issue, with the need for implicit or explicit coordination between applications, transport-, network-, and access-layer algorithms, there have been studies looking at the cross-layer congestion control (\textit{e.g.}, Kovacs \textit{et al.}~\cite{kovacs2008} and ETSI specialist task force work on cross-layer DCC~\cite{etsi14TR101612}). In terms of using awareness to adjust the parameters (power and rate) of congestion control algorithms, Gozalvez and Sepulcre propose OPRAM~\cite{gozalvez2008channel}, an opportunistic transmission power control algorithm that increases the transmit power of messages in critical situations (\textit{e.g.}, before intersections). However, in order to function properly, apart from precise location information, such as from GPS transmissions, OPRAM requires \textit{a priori}  knowledge about geographical regions that are accident-prone. Kloiber \textit{et al.}~\cite{kloiber2012dice,kloiber15} used awareness quality as a metric and employ a random transmit power for messages with a goal of reducing interference. Huang  \textit{et al.}~\cite{huang10} perform power and rate adaptation mechanisms independently, whereas the proposed mechanism, environment-and context-aware combined power and rate (ECPR), proactively considers the effect of power adaptation on rate adaptation and \textit{vice versa} such that it can adapt the mechanisms more efficiently at the next calculation step. Another difference is that Huang \textit{et al.}  perform rate adaptation based on potential tracking error resulting from the difference between actual and estimated states. This approach might be challenging to use in practice since it is hard to precisely obtain the actual state at each algorithm step. ECPR performs rate adaptation based on the channel utilization limit defined in the standards. Sepulcre \textit{et al.}~\bengi{\cite{sepulcre16}} proposes the integration of congestion and awareness control (INTERN), which adjusts transmit power based on the prevailing application context (target dissemination distance set by applications) alone, without considering the surrounding environment. Countless measurement studies have shown that the surroundings and vehicle traffic significantly affect the range, thus making it difficult to separate the target application range from the propagation environment restrictions. \bengi{Frigau \textit{et al.}~\cite{frigau15} control the transmission range using the transmission power as well as beacon generation range based on beacon reception rate. Nasiriani \textit{et al.}~\cite{nasiriani13} perform similar power control mechanism and combine it with rate control based on channel utilization. Jose \textit{et al.}~\cite{jose15} defines power adaptation as a joint Lagrangian optimization and rate adaptation. These approaches as well as~\cite{zemouri14} combine power and rate adaptation without their combined operation. However, the value that power control decides may cause a negative effect on message rate control mechanism, and vice versa.}

In order to enable safe and efficient cooperative vehicular communications, several technical challenges associated with VANETs include the following:
\begin{itemize}
\item Diverse interference caused by the other networks decreases vehicles communication efficiency.
\item Beacons and other messages cause increased overhead across the control channels. 
\item Dynamic environments need various control mechanisms. For example, if the message rate is fixed to a low value, this causes underutilization  in low density environments. Conversely, if the message rate is set to a high value, the vehicles may overload the channel in  high density traffic scenarios, thus causing collisions.
\item Each vehicle can have its own target awareness distance and target message rate. However, the current state-of-the-art does not provide  a practical solution for both distributed and coherent adaptation.
\end{itemize}

In this paper, we propose a transmit power control approach designed to achieve cooperative neighborhood awareness for vehicles, while the rate control is subsequently employed to utilize the available resources. Specifically, we propose an algorithm called ECPR (Environment- and Context-aware Combined Power and Rate Distributed Congestion Control for Vehicular Communication), which is a combined power and rate control DCC algorithm that aims to improve the cooperative awareness for challenging environments, while at the same time increasing the message rate when the environment and application requirements\footnote{We use the term ``application requirements" to encompass the effects that determine the rate and awareness requirements for a vehicle (\textit{e.g.}, speed, traffic conditions, and currently active application).} permits. To comply with target channel load/capacity requirements, ECPR employs an adaptive rate control algorithm. In this work, we use LIMERIC~\cite{bansal}, a state-of-the-art adaptive rate control algorithm, although other adaptive rate control algorithms could serve the same purpose. We performed simulations with ECPR in an experimentally validated simulation tool~\cite{boban14TVT} and showed that it can provide gains in terms of awareness or throughput in realistic propagation environments. \bengi{The proposed mechanism is briefly presented in ETSI 101 613~\cite{etsi101613}.}

Compared to current state-of-the-art, the main contributions of our work are:
\begin{itemize}
\item A practical algorithm to incorporate awareness -- a key building block for VANET applications -- as a core metric for congestion control in VANETs. ECPR proactively considers the effect of power adaptation on rate adaptation and \textit{vice versa}, so that it can adapt the mechanisms more efficiently at the next algorithm step.
\item By adjusting the transmit power based on the awareness criterion, we enable: i) congestion control adaptation to the dynamic propagation environment surrounding vehicles; and ii) effective adaptation of cooperative awareness range based on the application context, including requirements of different safety and non-safety applications, speed of vehicles, and different traffic conditions per direction.
\item By combining rate and awareness control, the proposed algorithm can achieve one of the following goals: i) improved channel utilization (in terms of the overall number of messages exchanged) for a given awareness rate; or ii) improved cooperative awareness for a given channel utilization; 
\end{itemize}

We perform extensive simulations including both realistic propagation and environment modeling (\textit{e.g.} large- and small-scale fading parameters, dynamic transitions between LOS and non-LOS links based on real building and vehicle locations) as well as realistic vehicle contexts (varying demand on both awareness by range and rate). We show that ECPR increases awareness by up to $20\%$ while keeping the channel load within reasonable bounds and interference at almost the same level. When the target awareness distance permits it, our proposed algorithm improves the average message rate by approximately 18\%, while keeping the target awareness. 

\begin{figure}[!t]
\centerline{\includegraphics[width=0.85\textwidth]{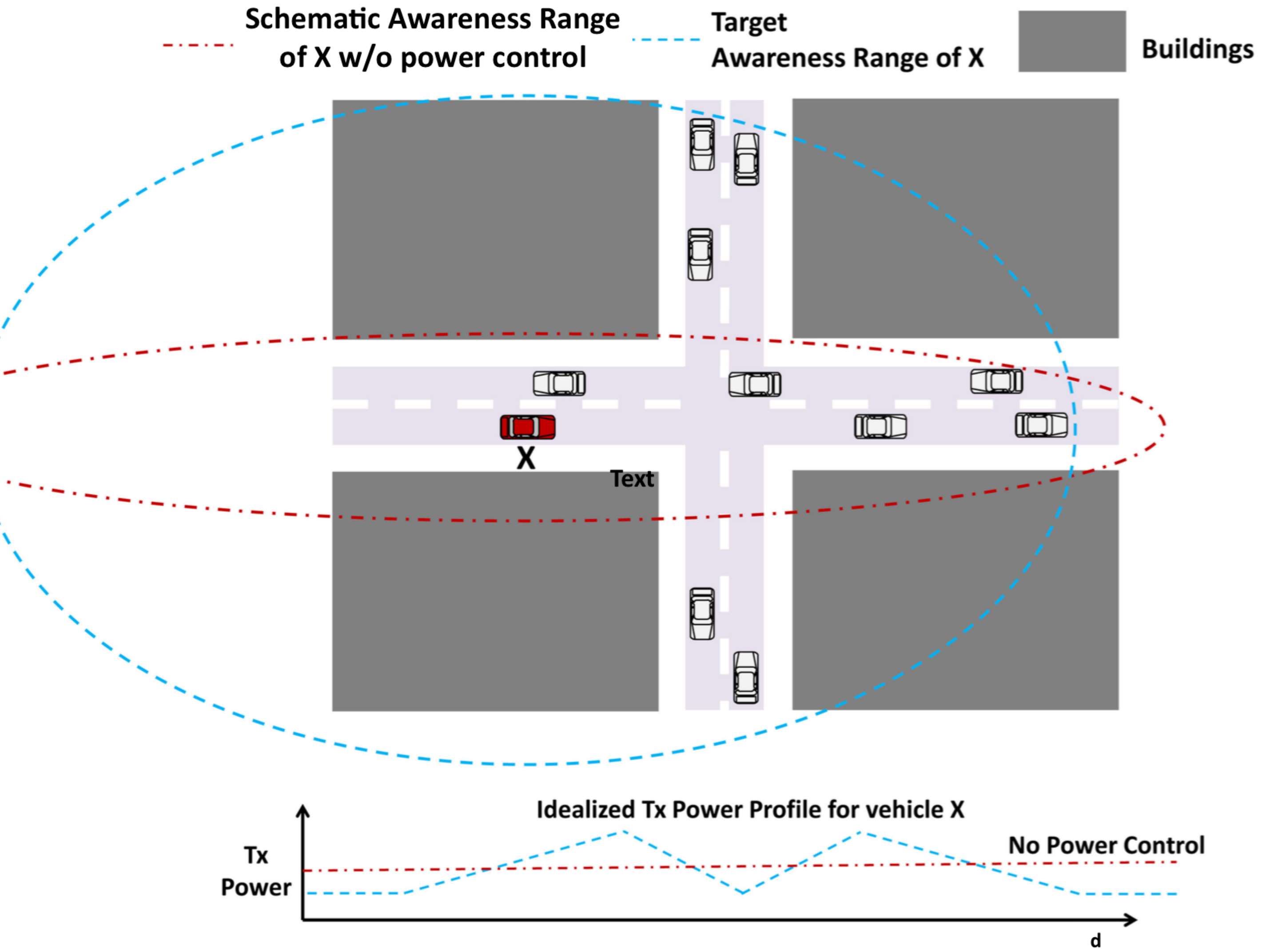}}
\caption{An example of how environment shapes the awareness range. Due to the particular environment layout, with buildings surrounding the intersection, if it is using fixed transmit power, vehicle X is likely to inform the vehicles on the same road of its existence, with a limited awareness of vehicles on the perpendicular road, up until X is in the intersection, at which point vehicles on both roads are likely to be aware of it. However, for active safety applications, awareness of vehicles on perpendicular road is more valuable than that on the same road, since the drivers of those vehicles cannot see vehicle X. Thus, for most VANET applications, \bengi{it is assumed that} the target awareness/communication range is a circular shape (or as circular as possible) of certain radius. Achieving such range in different environments requires power control. Lower part of the figure shows an idealized transmit power profile to adapt to the intersection environment for vehicle X as it travels through the intersection.} \label{SysArchUrban} 
\end{figure}

\begin{figure}[!t]
\centerline{\includegraphics[width=0.85\textwidth]{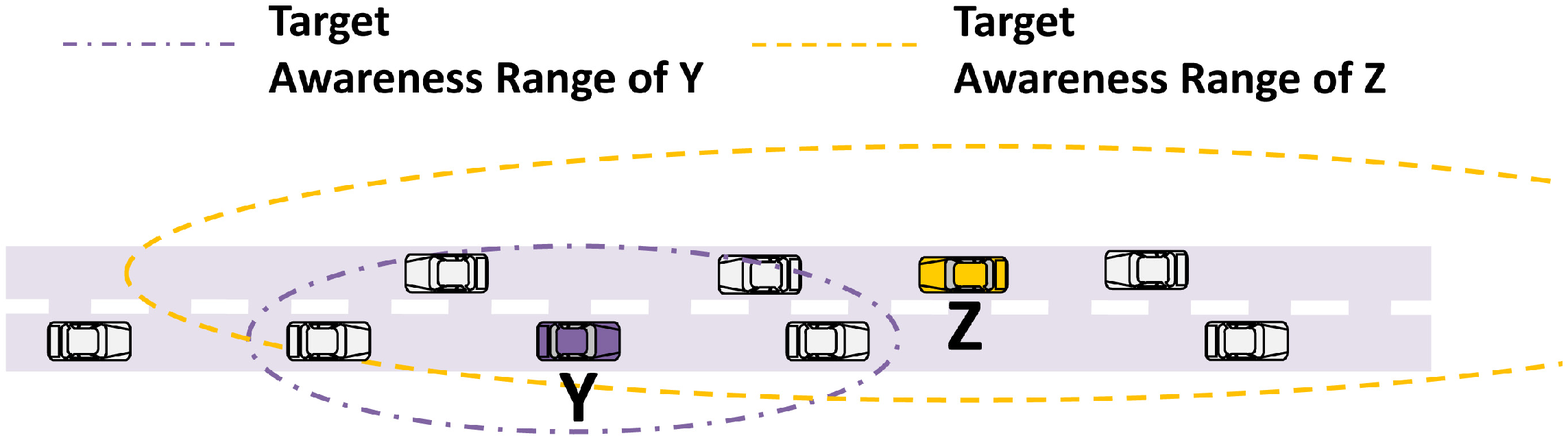}}
\caption{Depending on the application context, which includes the speed of the vehicle, traffic context and the type of currently active application~\cite{etsi10_2}, vehicles can have different target awareness ranges. For example, vehicle Y can be going at a lower speed than vehicle Z, in which case it might require smaller awareness range. Similarly, vehicle Z might be executing a safety-critical application (\textit{e.g.}, emergency vehicle notification), in which case it requires larger awareness range} \label{SysArchHway} 
\end{figure}

The rest of the paper is organized as follows: In Section~\ref{Network Architecture}, we describe the problem, provide several real-world scenarios, and define the metrics for evaluation of DCC algorithms. In Section~\ref{Algorithm}, we describe our proposed DCC approach. In Section~\ref{Experiment Results}, we discuss experiment results, and several concluding remarks are made in Section~\ref{Conclusion}. 
%
%
%
%
%
%
%
%
%
%
\section{Environment- and Application Context-aware Congestion Control}\label{Network Architecture}
The work presented in this paper aims at designing a novel DCC solution for V2V communication that can satisfy the target awareness levels for different application contexts in different realistic propagation environments. As noted earlier, cooperative awareness is vital for VANETs since many applications need to be aware of neighboring vehicles to trigger the correct type of action for avoidance of hazardous situations (\textit{e.g.}, accident prevention). To that end, in this section we discuss the main design goals for DCC algorithm and introduce metrics we use for evaluation of the algorithms.

\subsection{Design goals}

To obtain acceptable performance in terms of cooperative awareness, DCC algorithms need to take into account the following aspects: 
\begin{itemize}
\item Application context, determined by vehicular traffic conditions and application constraints, yields the requirements in terms of rate (amount of data) and communication and awareness range. Based on the application context, the DCC algorithm needs to distribute the available channel resources in a fair way (fair both in terms of achieved awareness and rate). 
\item Due to varying vehicular traffic density and mobility, the network topology is highly dynamic and depends on the time of day, type of road and other features~\cite{aygun14,viriyasitava15}. The DCC algorithm needs to be adaptive with respect to network dynamics at a rate higher than the rate of change of network. 
\item The propagation environment where vehicular communication occurs can be highly varying, even within a relatively small area. Environment characteristics of urban, suburban and rural areas create different challenges for congestion control and awareness~\cite{boban14vnc}. The environment creates effects similar on network topology to that of varying traffic density and mobility, albeit with geographically constrained dynamics. 
\item In addition to the effect of static objects near the road, surrounding vehicles also introduce significant variation in the reception probability and network topology.  Depending on vehicle size, a vehicle can completely block the communication between two other vehicles~\cite{abbas12}. Hence, a vehicle on a highway with dense traffic (\textit{e.g.}, morning rush hour) will have larger number of neighbors and a limited communication range due to the obstruction by surrounding vehicles; on the same highway during late of night, a vehicle will have fewer neighbors and an increased range. The DCC algorithm should be able to adapt to such variations. 
\item Electromagnetic emission regulations, limited channel resources, and potentially high number of communicating entities (including vehicles and roadside units) create practical limits on the ability to control the power and rate parameters.
\end{itemize}

Figure~\ref{SysArchUrban} shows how the physical environment affects the awareness range \bengi{\cite{sommer13}}, whereas  Figure~\ref{SysArchHway} shows how the application context requirements affect the target awareness range. In reality, there will exist numerous scenarios where the effects of the environment and application context will be combined, with the applications setting the awareness and rate requirements and the environment shaping the awareness range. Our goal in this study is to design a DCC solution that can efficiently support the functioning of safety and non-safety applications in diverse and dynamic VANET scenarios.
 
\subsection{Metrics}\label{subsec:Metrics}

\begin{figure}[t!]
\vspace{-50pt}
  \begin{center}
  \subfigure[\scriptsize Highway.]{\label{fig:NARFH}\includegraphics[height=0.3\textwidth]{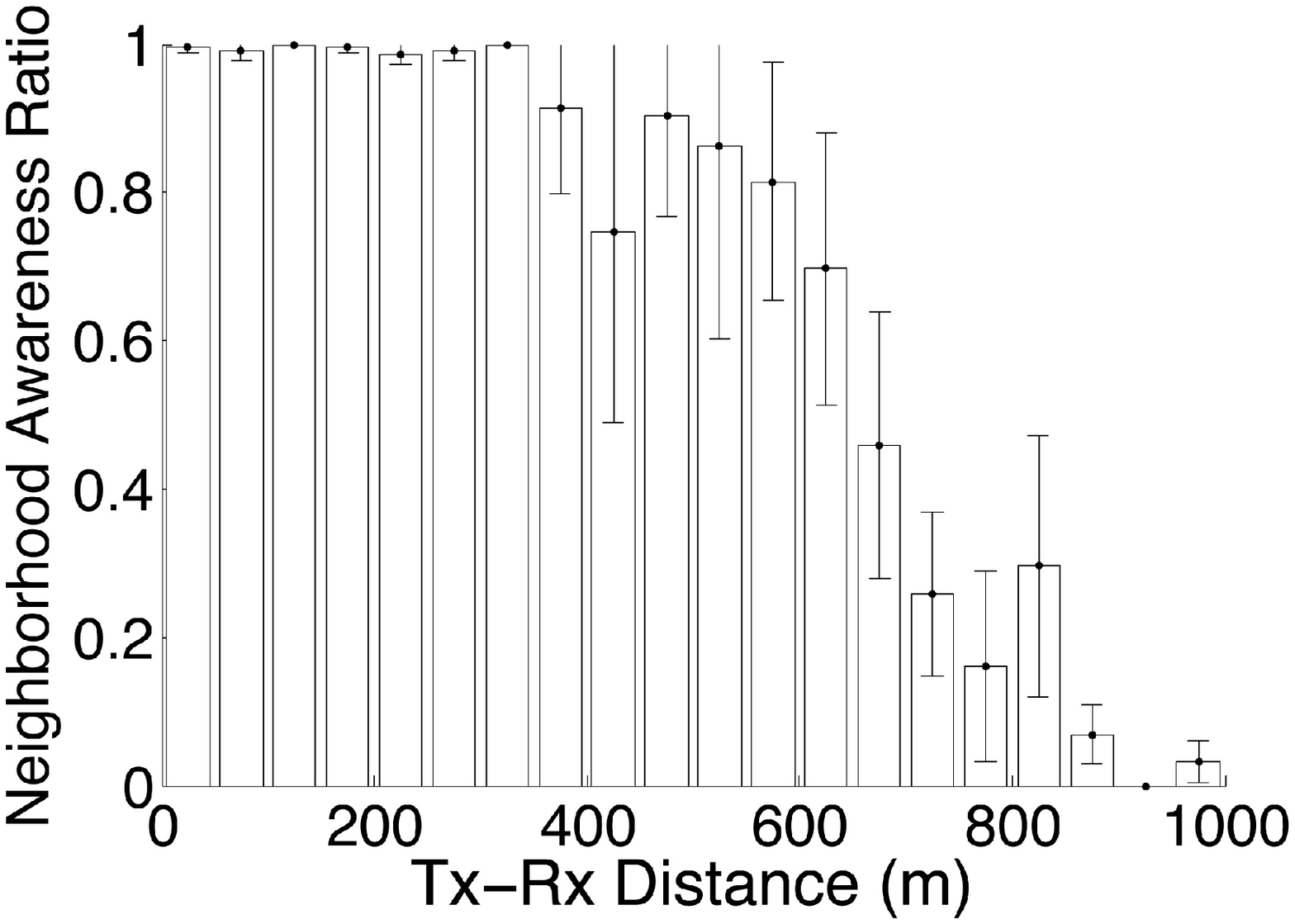}}\hspace{1mm}
   \subfigure[\scriptsize Urban.]{\label{fig:NARFU}\includegraphics[height=0.3\textwidth]{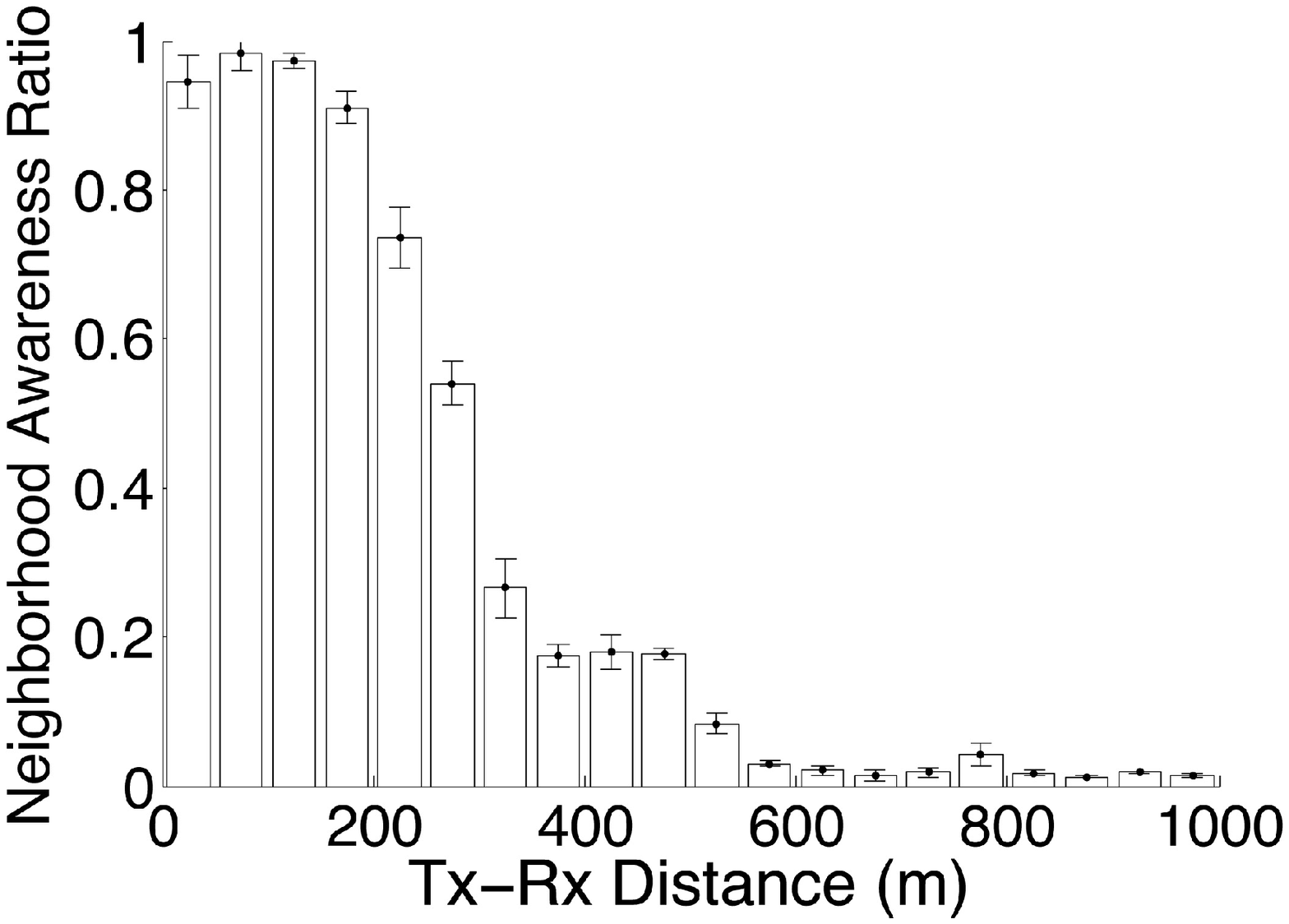}}\hspace{1mm}
     \caption{Measurements of NAR in Tampere, Finland. Measurements in both environments were collected using in the same measurement run based on the same vehicles, fixed transmit power, and 10 cooperative messages sent per second~\cite{boban14vnc}. } 
\label{fig:NARMeasurements}      
   \end{center}
\end{figure}

One of the main goals of cooperative awareness is to enable drivers/vehicles to enhance their knowledge of the environment in order to augment the information that they can obtain visually. To that end, cooperative message exchange mechanisms need to ensure that vehicles are aware of other relevant vehicles within the same geographical proximity, including those that are in non-LOS conditions. However, achieving this goal efficiently is a challenge since environments where vehicular communication occurs are quite diverse. For example, the transmit power required to send a message to a vehicle in an open environment (\textit{e.g.}, highway scenario) at a certain distance will likely be much lower than the power required to send a message to a vehicle at the same distance in a non-LOS environment (\textit{e.g.}, urban scenario) as shown in Figure~\ref{fig:NARMeasurements}.

To evaluate cooperative awareness in vehicular environments, we use two metrics introduced in previous work~\cite{boban14vnc}: Neighborhood Awareness Ratio (NAR) and Ratio of Neighbors Above Range (RNAR). For completeness, we define these metrics as follows:
\begin{itemize}
 \item\textbf{\textit{NAR}}: The proportion of vehicles in a specific range from which a message was received in a defined time interval. Formally, for vehicle $i$, range $r$, and time interval $t$, $NAR_{i,r,t} = \frac{ND_{i,r,t}}{NT_{i,r,t}}$, where $ND_{i,r,t}$ is the number of vehicles within $r$ around $i$ from which $i$ received a message in $t$ and $NT_{i,r,t}$ is the total number of vehicles within $r$ around $i$ in $t$ (we use $t$=1 second). This metric measures the ability of cooperative message exchange to fulfill its purpose: enable cooperative awareness. 
 \item \textbf{\textit{RNAR}}: For a vehicle $i$, range $r$, and time interval $t$, the ratio of neighbors that are above a certain distance from the observed vehicle $RNAR_{i,r,t} = \frac{NA_{i,r,t}}{N_{i,t}}$, where $NA_{i,r,t}$ is the number of vehicles above $r$ from which $i$ received a message in $t$ (again, we use $t$=1 second) and $N_{i,t}$ is the total number of vehicles from which $i$ received a message in $t$ (irrespective of $r$). This metric gives an indication of potentially unnecessary traffic overheard from distant neighbors (\textit{i.e.}, those that are not relevant for current application context). Once the technology is  deployed at a large scale (\textit{i.e.}, with communication equipment installed in most vehicles), such traffic will translate to unwanted interference.
\end{itemize}

In addition to NAR and RNAR, we also analyze the the performance of DCC in terms of the following metrics.
\begin{itemize}
\item \textbf{\textit{Average Message Rate}} shows the number of messages that a vehicle can transmit per second, averaged over all vehicles for a given second.
\item \textbf{\textit{Average Transmit Power}} shows the average transmit power messages that a vehicle transmits, averaged over all vehicles for a given second.
\item \textbf{\textit{Channel Busy Ratio (CBR)}} is defined as the proportion of channel time where the energy measured on the channel is above the Clear Channel Assessment (CCA) threshold.
\end{itemize}

\section{Proposed ECPR Algorithm}
\label{Algorithm}
In this section, we describe the proposed ECPR (Environment- and Con--text-aware Combined Power and Rate Distributed Congestion Control) algorithm. The goal of ECPR is to satisfy the requirements of target awareness levels for different application contexts in different realistic propagation environments, along with utilizing the available channel resources. Due to possibly different application contexts and environments, the vehicles will have different target awareness ranges and different target rates. To that end, ECPR uses power to control awareness range (distance) for the vehicles, whereas it uses rate to utilize the channel resources as allowed by the awareness requirements. In other words, ECPR attempts to satisfy the awareness requirements, at the same time maximizing the rate of messages through rate control. If the vehicles require low rates in order to not overload the channel, ECPR will set the transmit power of the vehicles to a maximum value. However, when the channel load increases (either due to higher rate requirements or due to an increased number of vehicles), ECPR is able to reduce the power in order to support such scenarios by considering  the awareness requirement. Below we explain how power and rate control components are implemented, along with the way they are combined to reach the abovementioned goals. 

\subsection{Power Adaptation for Awareness Control}
\label{Power Adaptation}
The power adaptation component of ECPR adapts the transmit power based on the current target awareness range set by the application context. ECPR is capable of adapting to dynamic scenarios with varying application contexts and in different environments without requiring \textit{explicit knowledge about the surroundings}, such as map information. To do so, it needs to estimate the channel \bengi{path loss} for all vehicles from which a message has been received the past time segment \textit{t}. Consequently, each vehicle requires knowledge of the transmit power level of the messages sent from each of its neighbors. The value of neighbor's transmit power information can be transmitted in the form of an integer value (e.g., between 0 and 33 dBm), which can be piggybacked in the transmitted messages (\textit{e.g.}, in cooperative awareness messages or in data packets).

To adjust the transmit power in order to meet the awareness requirement, ECPR use Path Loss Exponent (PLE) estimation. The transmit power adaptation algorithm is described as follows:

\begin{itemize}
\item \textbf{Define:} \textit{Ego vehicle:}The vehicle that is currently estimating its DCC parameters;\\
						\indent{\hspace{40pt}\textit{Neighbor:} Vehicle from which ego vehicle received a message within time segment $[t-1,t]$ \bengi{sec}} 
\item \textbf{Given:} Ego vehicles' transmit power at time $t$: $P_e^{Tx}(t)$; \\
\indent{\hspace{40pt}$i^{th}$ neighbor's transmit power at time $t$: $P_{i}^{Tx}(t)$, where $i = 1,..., N$ ($N:$ \bengi{Known number of neighbors within range});} \\
\indent{\hspace{40pt}Target awareness range of ego vehicle $r_e(t)$;} \\
\indent{\hspace{40pt}Target awareness percentage of ego vehicle within $r_e(t)$  \bengi{(Target NAR described in Section~\ref{subsec:Metrics})} : $TA_e(t)$}
\item For each received message, calculate $d_{ij}(t)$, distance between ego vehicle and $i^{th}$ neighbor at time $t$ when message $j$ was received
\item Select neighbors that are within target awareness range $r_e(t)$; select messages which are received from neighbors within $r_e(t)$
\item Compute $PLE_{ij}(t)$ (PLE for message $j$ from neighbor $i$) by using log-distance path loss as per~\cite{rappaport96}:
\bengi{
\begin{align}
PLE_{ij} (t) =\frac{PL_{ij} (t)}{10\log_{10}\left(\frac{4 \pi}{\lambda}d_{ij}(t)\right)},
\end{align}
where $\lambda$ the signal wavelength} and $PL_{ij}(t)$  is the path loss for message $j$ of neighbor $i$: 
\begin{align}
PL_{ij}(t) =P_{i}^{Tx}(t) - P_{ij}^{Rx}(t),
\end{align}
where $P_{i}^{Tx}(t)$ and $P_{ij}^{Rx}(t)$ are the transmit (Tx) of neighbor $i$ and receive (Rx) power of $j^{th}$ message from neighbor $i$, respectively.
\item Calculate ego's nodes transmit power required to reach $i^{th}$ neighbor for next time step, $P_{e\rightarrow{i}}^{Tx} (t+1)$, using  
$PLE_{ij} (t)$ and calculating the mean transmit power required for messages received from $i^{th}$ neighbor (with the mean over messages taken so as to counter the effects of fading):
\bengi{
\begin{align}
P_{e\rightarrow{i}}^{Tx} (t+1) = \frac{1}{m}\sum_{j=1}^m P_{ij}^{Rx} (t) + 10 PLE_{ij} (t) \log_{10}\left(\frac{4 \pi}{\lambda}r_e(t)\right).
\end{align}
}
\item Set ego node's transmit power for next time step ($t+1$) by considering the target awareness distance $r_e(t)$ and target awareness percentage $TA_e(t)$, provided as input of the application context. Sort the required transmit power to each neighbor and select $TA_e(t)$-th percentile transmit power: 
\begin{align}
& {P\_sorted}_e^{Tx} = sort_{i=1}^N(P_{e\rightarrow{i}}^{Tx} (t+1)),\\
&P_e^{Tx} (t+1) =  {P\_sorted}_e^{Tx} [round(TA_e(t)*N)].
\end{align}
\end{itemize}

Implicitly, by estimating the PLE from the received messages to adjust the transmit power, ECPR estimates what are the ``worst'' channels with \textit{all} vehicles within the awareness range $r_e$ (\textit{i.e.}, not only those from which a vehicle received messages correctly). By receiving messages from enough neighbors, ECPR gets an idea at what transmit power messages need to be sent at in order to reach the vehicles in $r_e$. In other words, by using PLE estimation, ECPR attempts to reach even those vehicles from which the ego vehicle has not yet received a message. \bengi{As long as the received power is higher than carrier sensing threshold, the transmit power at the next time step to the corresponding neighbor can be estimated. For the extreme cases such as very large path loss with a short distance, probably more than one neighbor will suffer from large path loss issue in the current environment. In that case, ECPR calculates Equation 4-5 and keep the transmit power high to reach the target awareness. The frame error level (less than $<5\%$) is neglected since there is not a significant impact on performance. It will be shown in Section~\ref{Experiment Results} that ECPR is a robust adaptation mechanism even in situations with significant MAC layer collisions.}

\subsection{Rate Adaptation}
\label{Rate Adaptation}
In this work, we employ the LInear MEssage Rate Integrated Control (LIMERIC) algorithm~\cite{bansal} to perform the rate adaptation aspect of ECPR due to its ability to converge to a fair and efficient channel utilization.\footnote{We note that ECPR is capable of performing combined adaptation for congestion control with other adaptive rate control algorithms.} LIMERIC takes the current channel busy ratio (CBR) and the current beacon rate as an input to the rate adaptation algorithm. The next beacon rate is adjusted to keep the current CBR under the threshold CBR, which is set to $0.6$ in this paper~\cite{etsi14TR101612}. The next message rate ($R_j(t)$) adaptation is done by Monte Carlo iteration at each ego node as defined below:
\begin{align}
R_j(t) = &(1-\alpha)R(t)(t-1) + sign(R_g- R_c(t-1))*\\ \nonumber
&min[X,\beta*|R_g-R_c(t-1)|],
\label{eq:Limeric}
\end{align}
where $R_c$ is the message rate, $\alpha$ and $\beta$ are the convergence parameters, and $R_g$ is target rate which satisfies the threshold CBR. For a detailed description of LIMERIC, we refer the reader to Bansal \textit{et al.}~\cite{bansal}. 

Recent measurement-based studies showed that message exchanges in vehicular environments are dominated by shadowing scenarios (\textit{i.e.}, obstruction by buildings, vehicles), where messages are both received and lost in bursts depending on the channel quality~\cite{wangsimulating14,boban14vnc}. This implies that sending fewer high-power messages in non-LOS scenarios have a better chance of creating awareness between vehicles than sending multiple successive messages at a lower transmit power. However, the current state-of-the-art with respect to DCC algorithms do not provision for making sure that the hard-to-reach vehicles are informed via cooperative awareness message exchange. Furthermore, depending on the speed of the vehicle, the type of traffic context (\textit{e.g.}, congested highway, busy or empty intersection) and the type of active application~\cite{etsi10_2}, target regions of interest (which directly translates into awareness range) can vary for different vehicles. Rate-control-only algorithms, which are proposed for the initial iteration of V2X systems~\cite{etsi14TR101612}, cannot accommodate for different awareness ranges.
\subsection{Combining power and rate control}
\begin{table}[t!]
\caption{Parameters used in the proposed algorithm} 
\footnotesize
\begin{center}
\begin{tabular}{ l | l }
\hline
\textbf{Parameter} & \textbf{Definition}\\
\hline
$t$ & Time (sec)\\ \hline
$r_e(t)$ & Target awareness range at time $t$ (m)\\\hline
 \multirow{2}{*}{$P_{i}^{Tx}$} & Transmit Power of j'th message from neighbor i\\
				   & within $r_e(t)$ (dBm)\\ \hline
$P_{ij}^{Rx}$ & Rx Power of j'th message from neighbor i within $r_e(t)$ (dBm)\\ \hline
 \multirow{2}{*}{$d_{ij}(t)$} & $i^{th}$ neighbor's distance within $r_e(t)$ at time $t$ \\
 									  &when receiving message j (m)\\ \hline
$DefaultTxPwr$ & Default transmit power (dBm)\\ \hline
$TA_e(t)$ &Target awareness of ego node at time $t$ (no unit)\\ \hline
$CBR(t)$ & Channel Busy Rate at time $t$ (no unit)\\ \hline
$lm_j$ &Length of the $j$'th message received by ego vehicle (byte/sec) \\ \hline
$C$ & Capacity of channel in terms of time (byte/sec)\\ \hline
$a = 0.1$, $b =1/150$ & LIMERIC parameters (see eq.~\ref{eq:Limeric}) (no unit)\\ \hline
$CBR_{Th}$ & Threshold CBR (no unit)\\ \hline
$\delta_A$ & Difference between target and actual awareness (no unit)\\ \hline
$eNAR(t)$ & Estimated Neighbor Awareness Ratio at time $t$ (no unit)\\ \hline
 \multirow{2}{*}{$\delta_R$} &The ratio of the difference between\\
			                             	& target and actual rate to target rate (no unit)\\  \hline
$TR(t)$ & Target message rate at time $t$ (Hz)\\  \hline
$BR(t)$ & Message rate at time $t$ (Hz) \\ \hline
$\gamma$ & Awareness/rate preference coefficient (no unit)\\  \hline
\end{tabular}
\end{center}
\label{tab:AlgoParameters}
\end{table}
\begin{algorithm}[h!]
\caption{Environment-Aware Combined Power and Rate Control for Vehicular Communication (ECPR) algorithm}\label{euclid}
\begin{algorithmic}[1]
\Statex
\State $PL_{ij}(t) =P_{i}^{Tx}(t) - P_{ij}^{Rx}(t)$
\State $PLE_{ij} (t) =\frac{ PL_{ij} (t)}{10\log_{10}\left(\bengi{\frac{4 \pi}{\lambda}} d_{ij}(t)\right)}$
\Statex
\If {$Neighbor_{e\rightarrow{i}}(t) \in Neighborhood_e(t-1)$}
\State $P_{e\rightarrow{i}}^{Tx} (t) =  \frac{1}{m}\sum_{j=1}^m P_{ij}^{Rx} (t) + 10 PLE_{ij} (t) \log_{10}\left(\bengi{\frac{4 \pi}{\lambda}}r_e(t)\right)$\\ 
\Else
\State {$P_{e\rightarrow{i}}^{Tx}(t) \gets DefaultTxPwr$\\
\EndIf
\State ${P_{sorted}}_e^{Tx} = sort_{\forall i,j\in N}(P_{e\rightarrow{i}}^{Tx} (t+1))$
\State $P_e^{Tx} (t+1) =  {P_{sorted}}_e^{Tx} [round(TA_e(t)*N)]$
\Statex
\State $CBR(t) = \sum_{j=1}^n {lm_j}/C$
\State $BR(t+1) = (1-a)BR(t) + sign(CBR_{Th}- CBR(t))*min[X, b(CBR_{Th}- CBR(t))]$
\Statex
\State $\delta_A = TA_e(t)-eNAR(t)$
\State $\delta_R = \frac{TR(t)-BR(t)}{TR(t)}$ 
\If{$CBR(t)<CBR_{Th}$}
\State Apply $P_e^{Tx} (t+1)$
\Else
\If{$P_e^{Tx} (t+1) \leq P_e^{Tx} (t)$}
\State Apply $P_e^{Tx} (t+1)$
\ElsIf{$\delta_A \geq \gamma\delta_R$}
\State Apply $P_e^{Tx} (t+1)$
\Else
\State $P_e^{Tx} (t+1) \leftarrow{P_e^{Tx} (t)}$
\EndIf
\EndIf
}
\end{algorithmic}
\end{algorithm}

Algorithm~\ref{euclid} describes the steps of the ECPR algorithm, whereas Table~\ref{tab:AlgoParameters} summarizes the parameters used by ECPR. The proposed combined control algorithm adapts the next transmission power based on the current path loss ($PL_{ij}(t)$) and path loss exponent ($PLE_{ij}(t)$) for each message ($j$) received from the neighbors (See Alg.~1: Line~1-2). If the neighbor $i$ was already ego node's neighbor in the previous time step, the algorithm assigns the required transmit power to this neighbor based on the current $PL_{ij}(t)$, $PLE_{ij}(t)$, and target awareness range. Conversely, if this vehicle was not a neighbor to the ego node in the previous time step, a default value (e.g., 10~dBm or 23~dBm in our simulations) is used as needed in order for the transmission power to reach this neighbor. By using the default transmit power value, the ego node increases the probability of being heard by those nodes for which it does not know what kind of power is needed to reach them (See Alg.~1: Line~3-6). Once the ego node has the transmission power information it needs to reach each of the neighbors, it sorts these values from the least to the most. The next transmission power level of the ego node is chosen by considering the target awareness percentage. In other words, the smallest value that covers $TA\%$ for all neighbors is chosen as the next transmit power (See Alg.~1: Line~8-9). In terms of rate adaptation, ECPR adapts the rate by using the current message rate and channel load (\textit{i.e.} CBR).  The ratio of the messages received divided by the channel capacity is defined as the CBR (See Alg.~1: Line~10-11) - this is in line with the standardized CBR calculation approaches~\cite{etsi14TR101612}.

Furthermore, as Algorithm~\ref{euclid} shows, the transmit power control takes into account the channel load (CBR), such that the transmit power is not increased if the CBR threshold is reached. The power control algorithm interacts with the rate control, such that the power and rate control ``share the load'' in case of high CBR: the relationship between the target and current beacon rate BR and the current and target awareness determines whether or not the transmit power will be changed (either increased or reduced). The value of coefficient $\gamma$ determines whether awareness or rate control is prioritized (In this study, we use the same weight for the awareness and rate: $\gamma$=1). Furthermore, in the case of high CBR, ECPR prevents a significant increase of the channel load that could be caused by the application context suddenly increasing the target awareness range $r_e$. However, we note that safety-critical messages generated due to hazardous events are going to be sent at a high power and rate that are not governed by the DCC algorithm. Therefore, controlling the power and rate of cooperative messages will not affect safety-critical messages (See Alg.~1: Line~10-11). For clarity, Table~\ref{tab:States} shows the transmit power control actions undertaken by ECPR depending on the channel load (CBR), awareness, and rate. 

\begin{table}[t!]
\centering
\footnotesize
\caption{States that affect transmit power adaptation} 
\vspace{5pt}
\begin{tabular}{l c c c c}
\hline
\textbf{State} &\textbf{CBR} &  \textbf{Awareness} & \textbf{Rate} &  \multirow{3}{*}{\textbf{Transmit Power at \textit{t+1}}}\\
 											&	\textbf{vs.}				 & \textbf{vs.} 						& \textbf{vs.}				&\\
 											&\textbf{Target}			 & \textbf{Target} 					&\textbf{Target}			& \\ \hline
1 & $<$  & $<$  	& $=$  	& Apply $P_e^{Tx} (t+1)$\\
2 & $<$  & $\geq$  	& $=$  	& Apply $P_e^{Tx} (t+1)$ \\
3 & $<$  & $<$  	& $<$  	& Apply $P_e^{Tx} (t+1)$ \\ 
4 & $<$  & $\geq$  	& $<$  	& Apply $P_e^{Tx} (t+1)$ if $\leq$ $P_e^{Tx}(t)$\\
5 & $>$  & $<$  	& $=$  	& Apply $P_e^{Tx} (t+1)$ if $\leq$ $P_e^{Tx}(t)$ OR $\delta_A \geq \gamma\delta_R$\\
6 & $>$  & $\geq$  	& $=$  	& Apply $P_e^{Tx} (t+1)$ if $\leq$ $P_e^{Tx}(t)$\\
7 & $>$  & $<$  	& $<$  	& Apply $P_e^{Tx} (t+1)$ if $\leq$ $P_e^{Tx}(t)$ OR $\delta_A \geq \gamma\delta_R$ \\
8 & $>$  & $\geq$  	& $<$  	& Apply $P_e^{Tx} (t+1)$ if $\leq$ $P_e^{Tx}(t)$\\
\end{tabular}
\label{tab:States}
\end{table}

The awareness metric measures the awareness of \textit{neighboring vehicles} about the ego vehicle, thus it can be estimated at ego vehicle \textit{locally} by using the channel loss to each neighbor and the transmit power that will be used at the ego vehicle at $t+1$. Since obtaining the NAR metric from a receiver's perspective as defined in Section~\ref{subsec:Metrics} would require a vehicle to know about all vehicles within $r$ (in which case, by design, its NAR for $r$ would be $1$), we define the estimated NAR (eNAR) from transmitter's perspective as follows:
\begin{align}
eNAR_{r}(t) = \frac{ND'_{r}(t)}{N_{r}(t)},
\end{align}
where $N_{r}(t)$ is the number of vehicles within $r$ at time $t$ which ego vehicle \textit{detected}  (\textit{i.e.}, received a cooperative message from), and $ND'_{r}(t)$ is the estimated number vehicles in $N_{r}(t)$ that detected the ego vehicle, calculated as:
\begin{align}
ND'_{r}(t) =  \bengi{\epsilon \cdot}\sum_{i=1}^N{I(P_e^{Tx}(t-1)+PL_{e\rightarrow{i}}^{Tx}(t-1)>P_{Th}^{Rx})},
\label{ndt}
\end{align}
where $I$ is the indicator function, $PL_{e\rightarrow{i}}^{Tx}(t-1)$ is the channel loss from ego vehicle to neighbor $i$, and $P_{Th}^{Rx}$ is the receiver sensitivity threshold. Effectively, the ego vehicle uses the channel reciprocity theorem ($PL_{e\leftarrow{i}}^{Tx}=PL_{e\rightarrow{i}}^{Tx}$)~\cite{rappaport96} to estimate the proportion of its neighbors that were able to receive cooperative messages from it in the previous time step. \bengi{The estimation error for number of neighbors is defined as $\epsilon$ and is set to $[-10,10] \%$. It is possible that a comparatively high power signal is lost due to strong interference (although not too frequently, since CSMA/CA mechanism and congestion control mechanism are in place). Hence, Equation~(\ref{ndt}) can introduce false positive cases which lead to an inaccurate number of neighbors.} 

At low densities, when vehicles have a small number of neighbors, the eNAR estimate can be incorrect because of a small number of data points it needs to work with. However, in low density cases, vehicles will almost always be able to achieve the maximum rate and awareness, since the channel load at low densities will be low. Therefore, knowing the correct eNAR is not necessary. As the network density increases and vehicles start having more neighbors and they have a larger number of data points to work with (\textit{e.g.}, $100$ instead of $10$ neighbors), which makes the eNAR estimate more accurate.

\section{Simulation Setup}\label{Experiment Results}
\begin{figure*}[t!]
  \begin{center}
  \subfigure[Highway Scenario]{\label{highwayRegion}\includegraphics[height=0.35\textwidth]{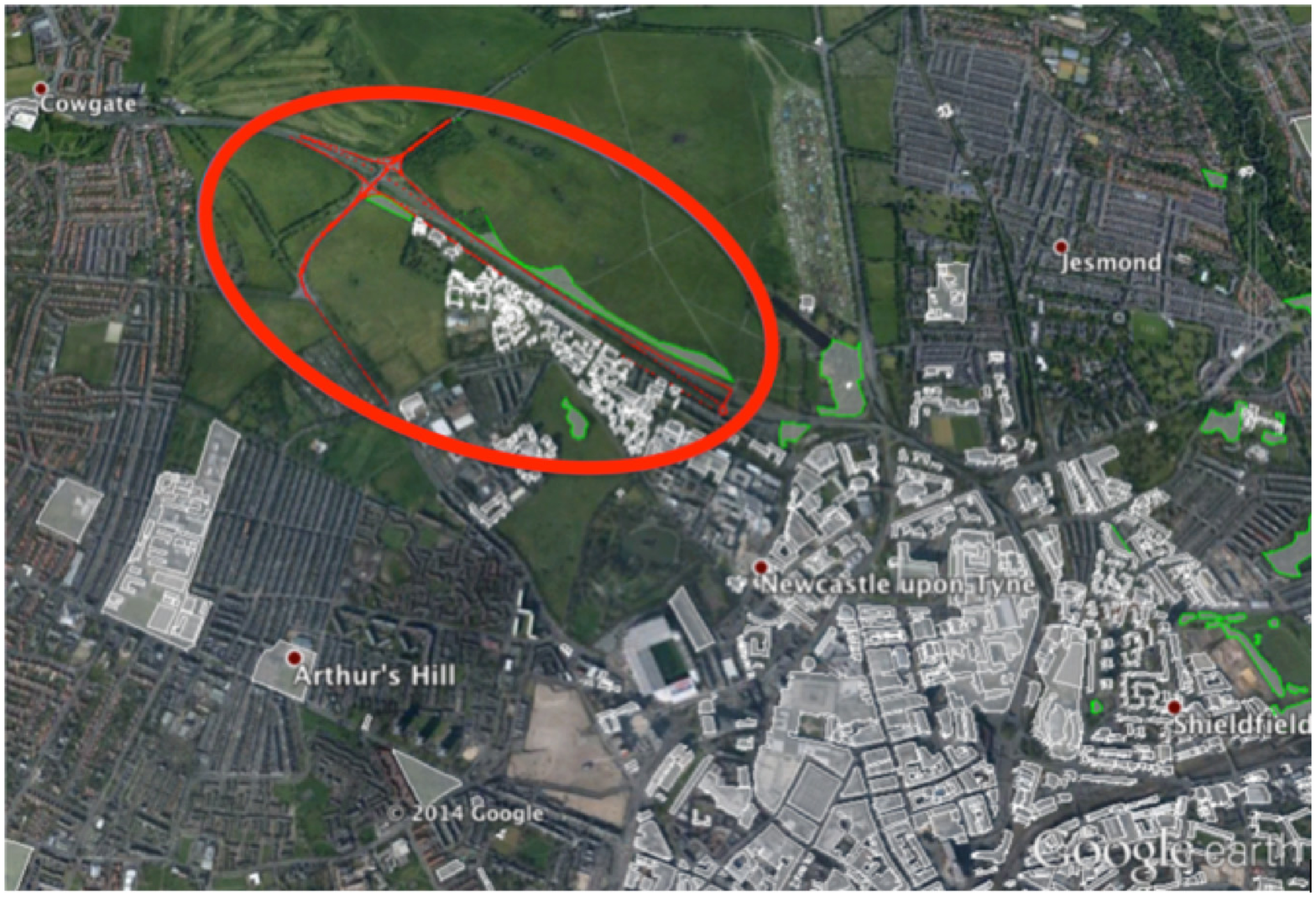}}\hspace{1mm}
   \subfigure[Urban Scenario]{\label{urbanRegion}\includegraphics[height=0.35\textwidth]{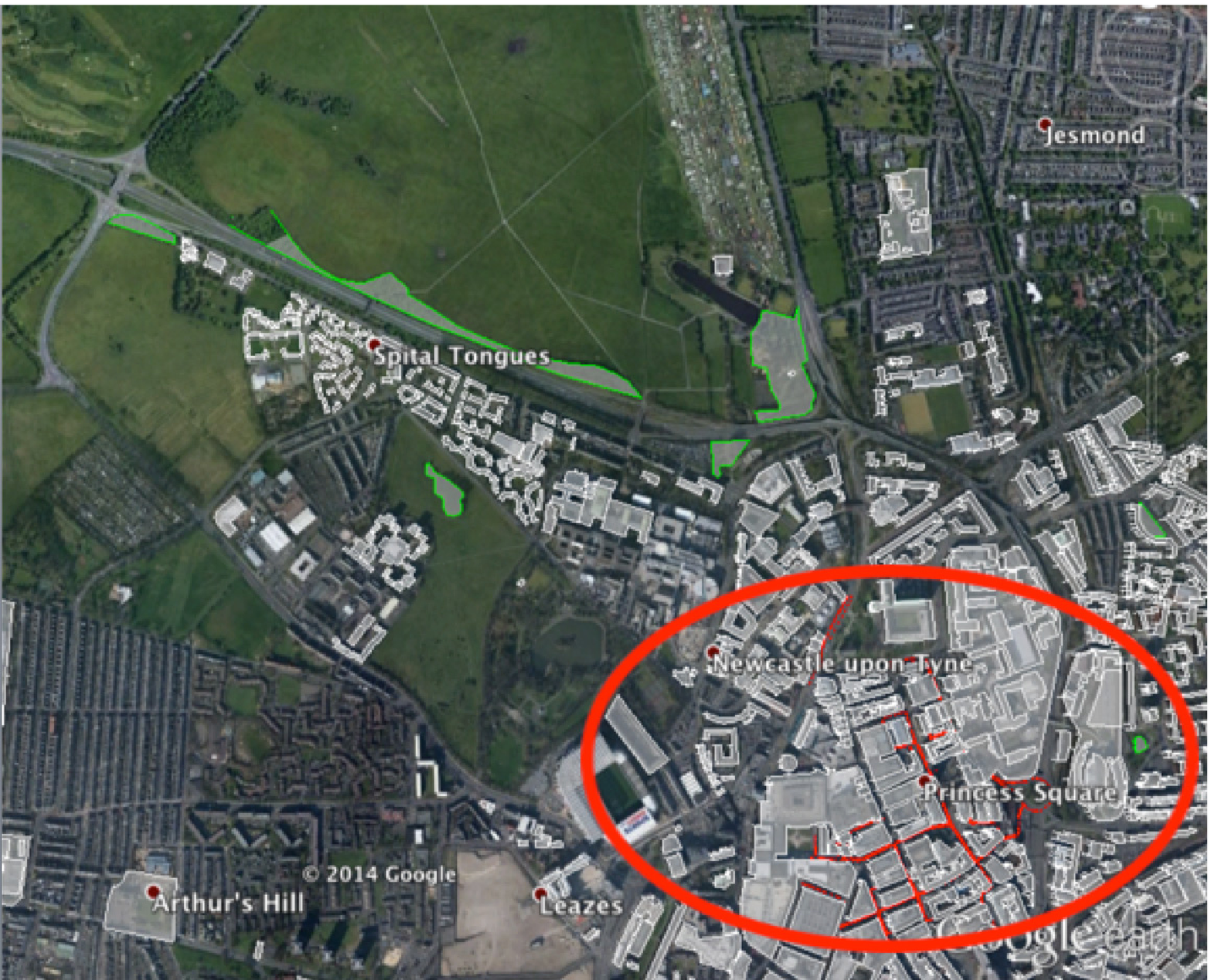}}\hspace{1mm}
     \caption{Regions used for highway and urban simulations (circled) on the topology of Newcastle, UK. Both regions have an area of approximately $1 km^2$. White outlines represent buildings that were incorporated in simulations for realistic propagation modeling.}  
      \label{region}
   \end{center}
\end{figure*}

To evaluate the performance of ECPR, we implemented it in the GEMV$^2$ V2V propagation simulator~\cite{boban14TVT}. GEMV$^2$ is a computationally efficient propagation model for V2V communications, which explicitly accounts for surrounding objects (\textit{e.g.}, buildings, foliage and vehicles~\cite{boban11}). The model considers different V2V links types (LOS, non-LOS due to static objects, non-LOS due to vehicles) depending on the LOS conditions between the transmitter and receiver to deterministically calculate large-scale signal variations. Additionally, GEMV$^2$ determines small-scale signal variations stochastically using a simple geometry-based model that takes into account the surrounding static and mobile objects (specifically, their number and size). By implementing ECPR in GEMV$^2$, we are able to show how it behaves in realistic propagation conditions, including varying LOS that affects the path loss and highly dynamic network topology changes caused by transition between environments (\textit{e.g.}, a vehicle on a road with low vehicular density moving to a high-density intersection). 

In terms of parameters, the time step used for the ECPR time step duration was set to $200$ $ms$. For a given target range $r$, we use a target awareness $TA=85\%$. We use omni-directional antennas on the vehicle roof and evaluate the DCC performance on a single channel. We set the maximum transmit power to $23~dBm$ and the maximum beacon rate to $10~Hz$. We used the performance metrics described in Section~\ref{subsec:Metrics}. 

To give a physical perspective to the parameters relevant for ECPR, the typical values for awareness range $r$ are from $20$ to $500~m$, depending on application context; similarly, target awareness within $r$, $TA$, will be dependent on the application context and can range from \textit{e.g.}, 50\% to 100\%; $P_e^{Tx}$ is usually limited from $0$ to $23~dBm$ in radios used for V2V communication, whereas the message rate $BR$ is usually set between $1$ and $10~Hz$ for cooperative messages~\cite{etsi14TR101612}. Communication parameters considered in this paper are summarized in Table~\ref{tab:parameters}.

Since the goal of this study is to show the feasibility of environment- and context-aware DCC control by leveraging the benefits of both power and rate adaptation, we choose to compare the proposed ECPR algorithm with LIMERIC (rate-only DCC algorithm), the power-control only component of ECPR, and a scenario without DCC (\textit{i.e.}, messages are set with fixed rate and power irrespective of the channel conditions). 

\begin{table}[t!]
\centering
\small
\caption{System Parameters}
\begin{tabular}{l c}
\hline
\textbf{Parameter} & \textbf{Value}\\ \hline
Carrier sense threshold [dBm]& -90\\  
Data rate [Mbps]& 6\\ 
Measurement period [ms]&200\\
Min. and Max.packet transmission frequency [Hz]&1 and 10 \\
Min. and Max. transmission power [dBm]&0 and 23 \\
Min. and Max. awareness range [m]&20 and 500 \\
Target neighbor awareness ratio& 85\%\\
Threshold Channel Busy Ratio& 60\%\\
\end{tabular}
\label{tab:parameters}
\end{table}

\subsection{Simulated environments}
One of the most challenging scenarios for DCC algorithms is to ensure they properly function in any kind of environment. To that end, we perform simulations using the city of \textit{Newcastle upon Tyne, England} as shown in Figure~\ref{region}. The region around \textit{A167} is chosen for the highway scenario. A part of the city grid around \textit{Princess Square} is used to simulate an urban area. We used $1$ $km^2$ area and $500$ vehicles for both the highway and urban simulations. Vehicular mobility is generated using SUMO~\cite{SUMO2012}, whereas OpenStreetMap~\cite{openstreetmap} is used to obtain the outlines of buildings and foliage for accurate propagation modeling. 

\subsection{Application Context: Varying Target Rate and Target Awareness Distance} 

As shown in Figure~\ref{SysArchHway}, depending on the application context, different vehicles can have different awareness range and rate requirements at the same time. To test ECPR with varying awareness range and rates, we perform four types of tests described in Table~\ref{tests}. In Test 1, each vehicle's target awareness range is set to $90~m$ and target beacon rate is $10~Hz$. In Test 2, the target awareness distance is 90 m and target beacon rate is different for all ego nodes. The target rate is chosen uniformly across an interval of $[5,10]~Hz$. In Test 3 and 4, the target awareness distances are selected uniformly at random. 
\begin{table}[t!]
\centering
\small
\caption{Tests defined with different target awareness range and message rate combinations to stress-test ECPR} 
\vspace{2pt}
\begin{tabular}{c c c}
\hline
 & \textbf{Target Awareness Range} &  \textbf{Target Message Rate} \\ \hline
\textbf{Test 1} & Same for all vehicles (90~m)  												 & Same for all nodes (10 Hz) \\ \hline
 \multirow{2}{*}{\textbf{Test 2}} &  \multirow{2}{*}{Same for all vehicles (90~m)}  & Uniformly distributed \\
 											& 																	 &between 5 and 10 Hz \\  \hline
 \multirow{2}{*}{\textbf{Test 3}} & Chosen randomly from set    						 &  \multirow{2}{*}{Same for all nodes (10 Hz)} \\
 											& S = [30, 60, 90, 120, 150, 180~m]					 &								\\  \hline
\multirow{2}{*}{\textbf{Test 4}} & Chosen randomly from set 							 &  Uniformly distributed \\
											&S = [30, 60, 90, 120, 150, 180~m]  			      &between 5 and 10 Hz\\  \hline
\end{tabular}
\label{tests}
\end{table}

\section{Results}
\subsection{Comparison of ECPR with LIMERIC, power-only algorithm, and no DCC}
In this subsection, we compare the performance of ECPR relative to LIMERIC (rate-only algorithm), the power-control only component of ECPR (described in Section~\ref{Power Adaptation}), and a scenario without DCC. To obtain a fair comparison, we use only Test 1 from Table~\ref{tests} (\textit{i.e.}, same awareness range and rate requirements for all vehicles). We perform simulations with different default transmit power settings: these affect the initial power levels for radios employed in the ECPR and power-only adaptation scenarios, whereas for no DCC and rate-only DCC scenarios the default power is used throughout the simulation. 

\begin{figure*}[th!]
  \begin{center}
     \subfigure[Neighbor awareness vs. transmission distance]{\label{narUrban150TxPwr10}\includegraphics[width=0.49\textwidth]{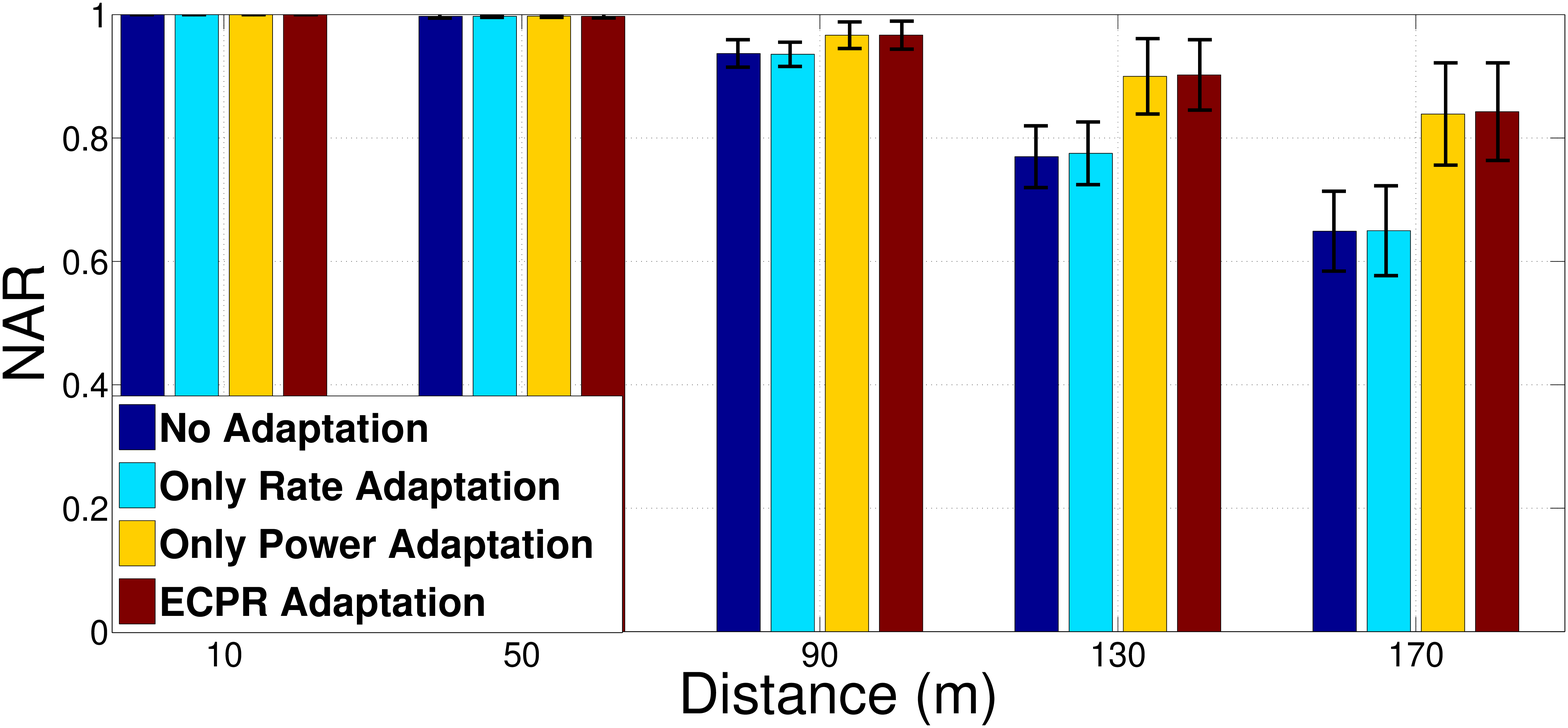}}\hspace{1mm}
\subfigure[Neighbors above range causes unwanted interference vs. transmission distance]{\label{rnarUrban150TxPwr10}\includegraphics[width=0.49\textwidth]{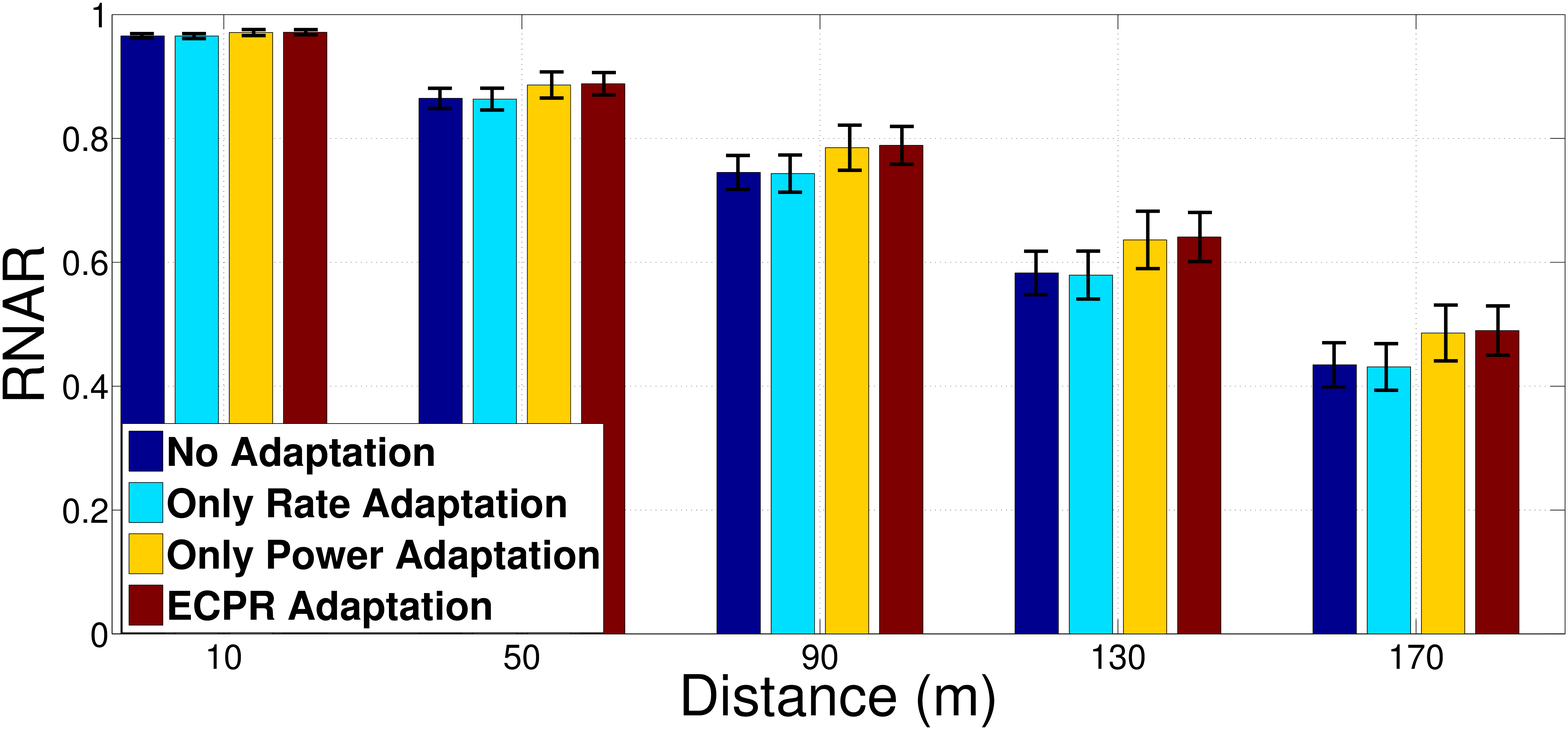}}\hspace{1mm}  
  \subfigure[Message  rate]{\label{rateUrban150TxPwr10}\includegraphics[width=0.49\textwidth]{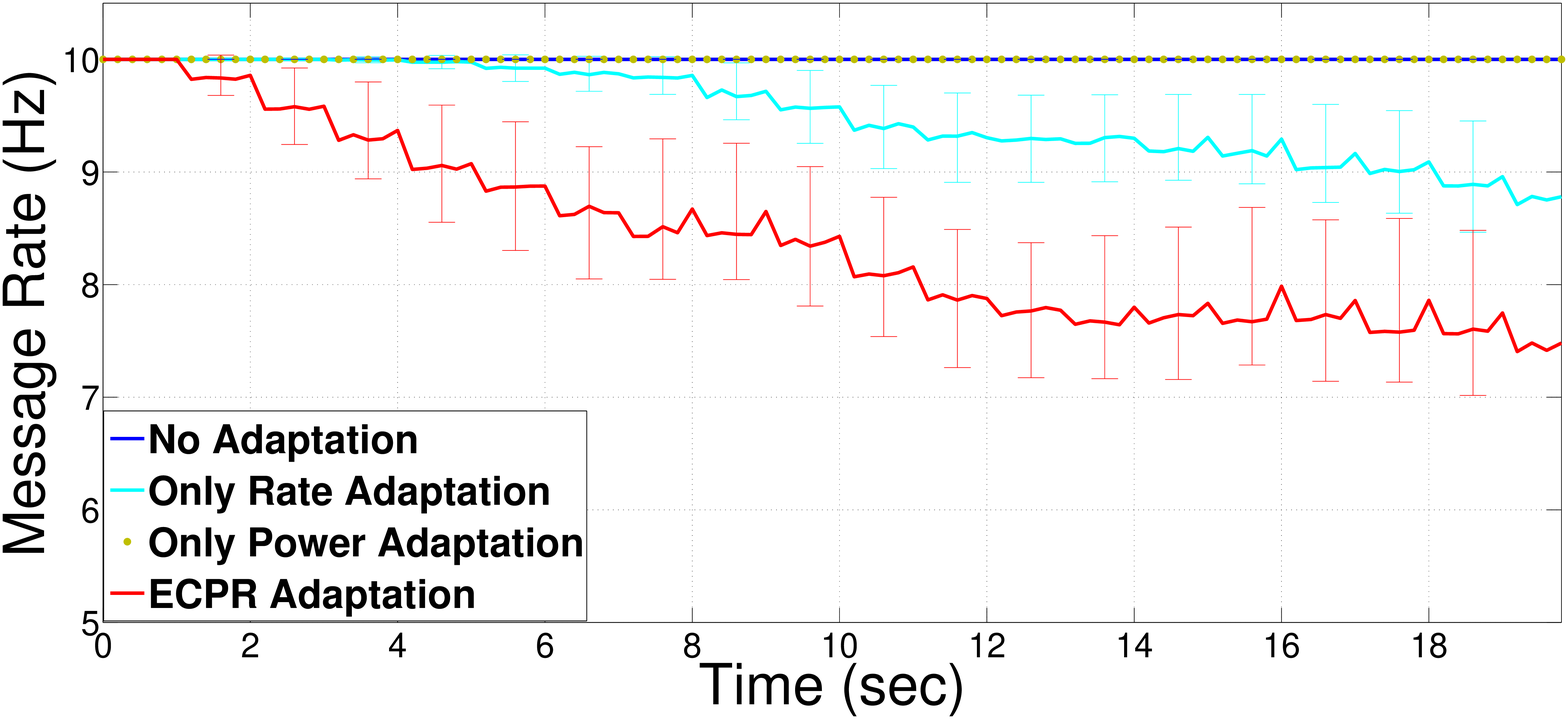}}\hspace{1mm}
 \subfigure[Transmit power]{\label{powerUrban150TxPwr10}\includegraphics[width=0.49\textwidth]{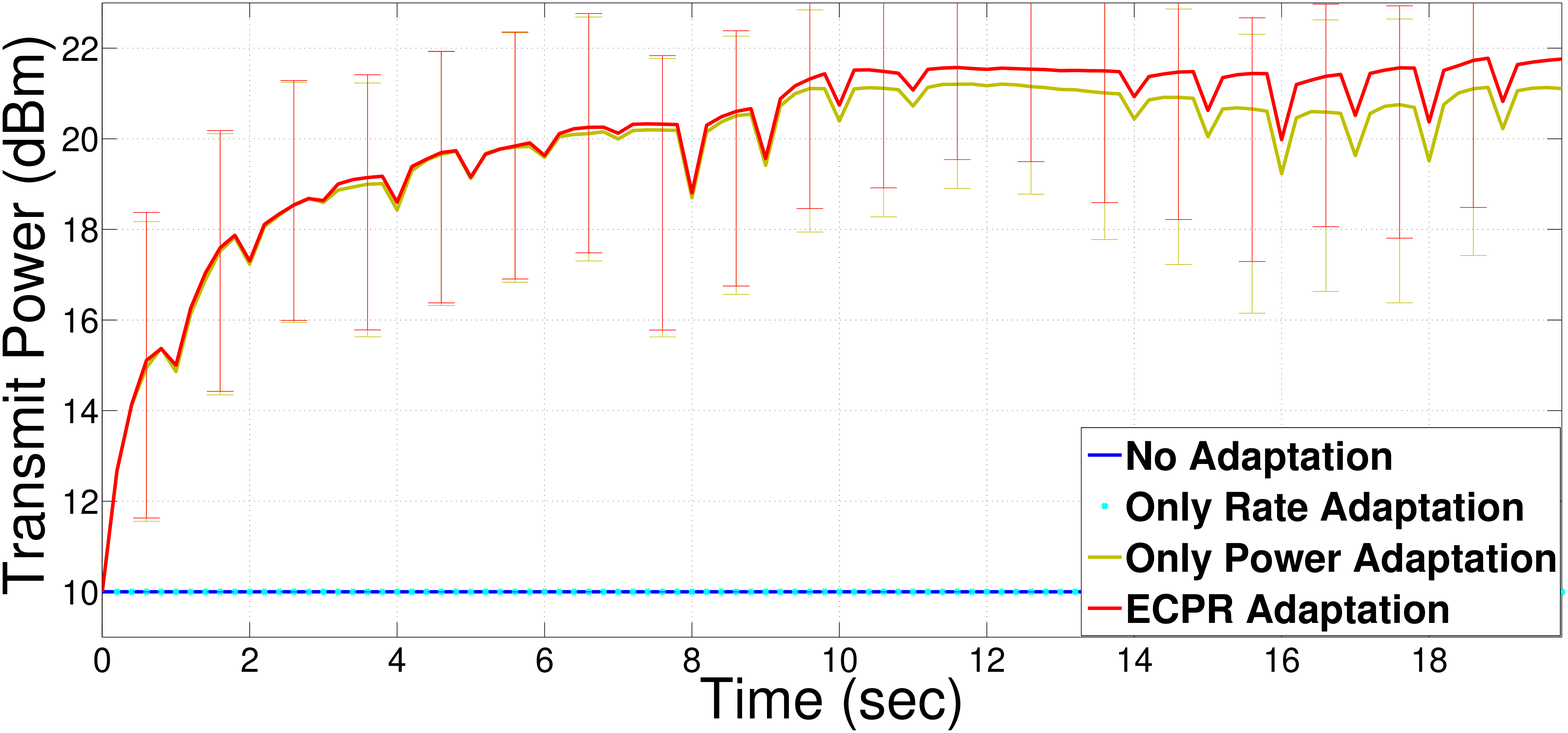}}\hspace{1mm}
 \subfigure[Channel busy ratio (CBR)]{\label{ cbrUrban150TxPwr10}\includegraphics[width=0.49\textwidth]{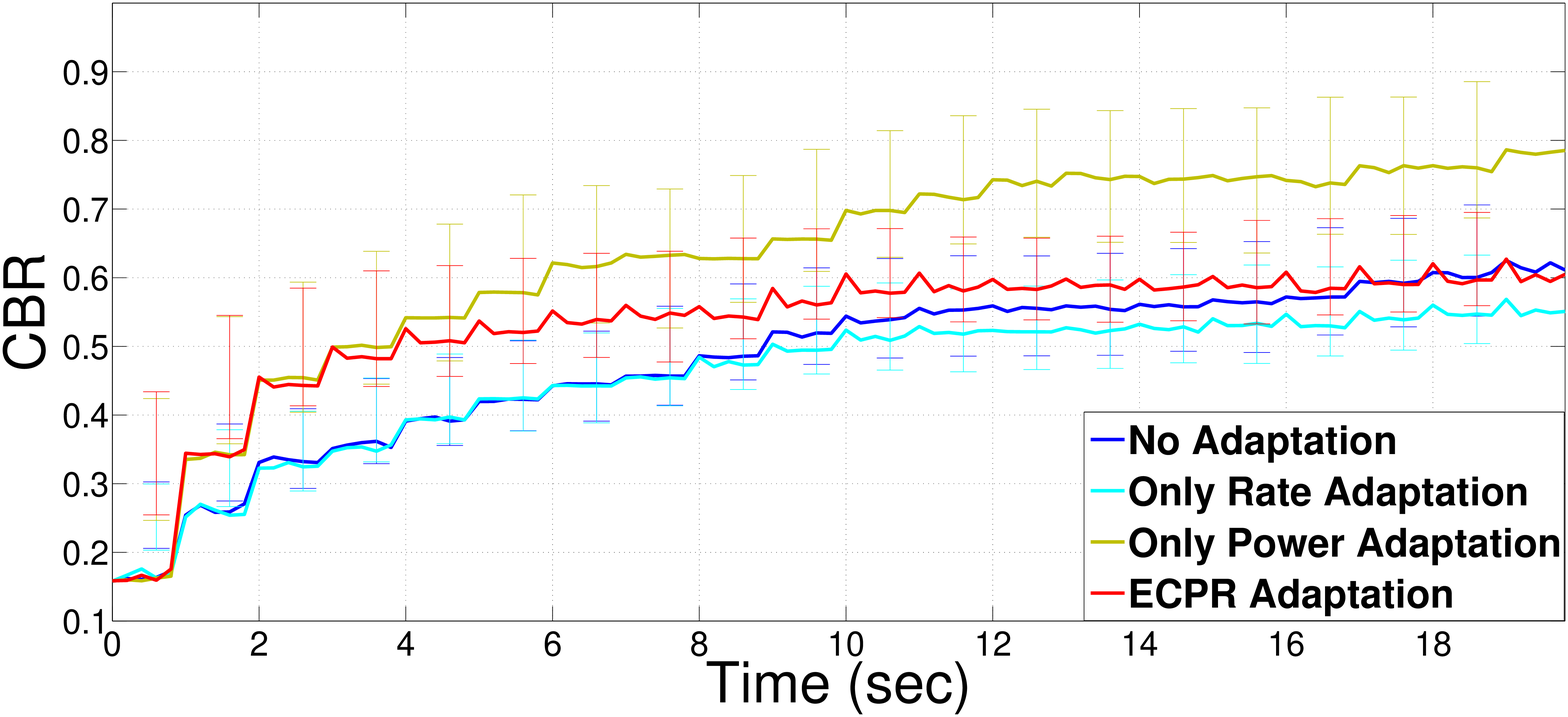}}\hspace{1mm}
     \caption{Target Awareness $85\%$, Target Awareness Distance = 150m, default Tx Power = 10 dBm. Urban Scenario. Power-only algorithm achieves awareness (NAR) comparable to ECPR; however, due to it not taking channel load (CBR) into account, it \bengi{exceeds} the target CBR. 
     }  
      \label{powerRateDiffAdp150}
   \end{center}
\end{figure*}

\begin{figure*}[t!]
  \begin{center}
     \subfigure[NAR]{\label{narUrban50TxPwr23}\includegraphics[width=0.49\textwidth]{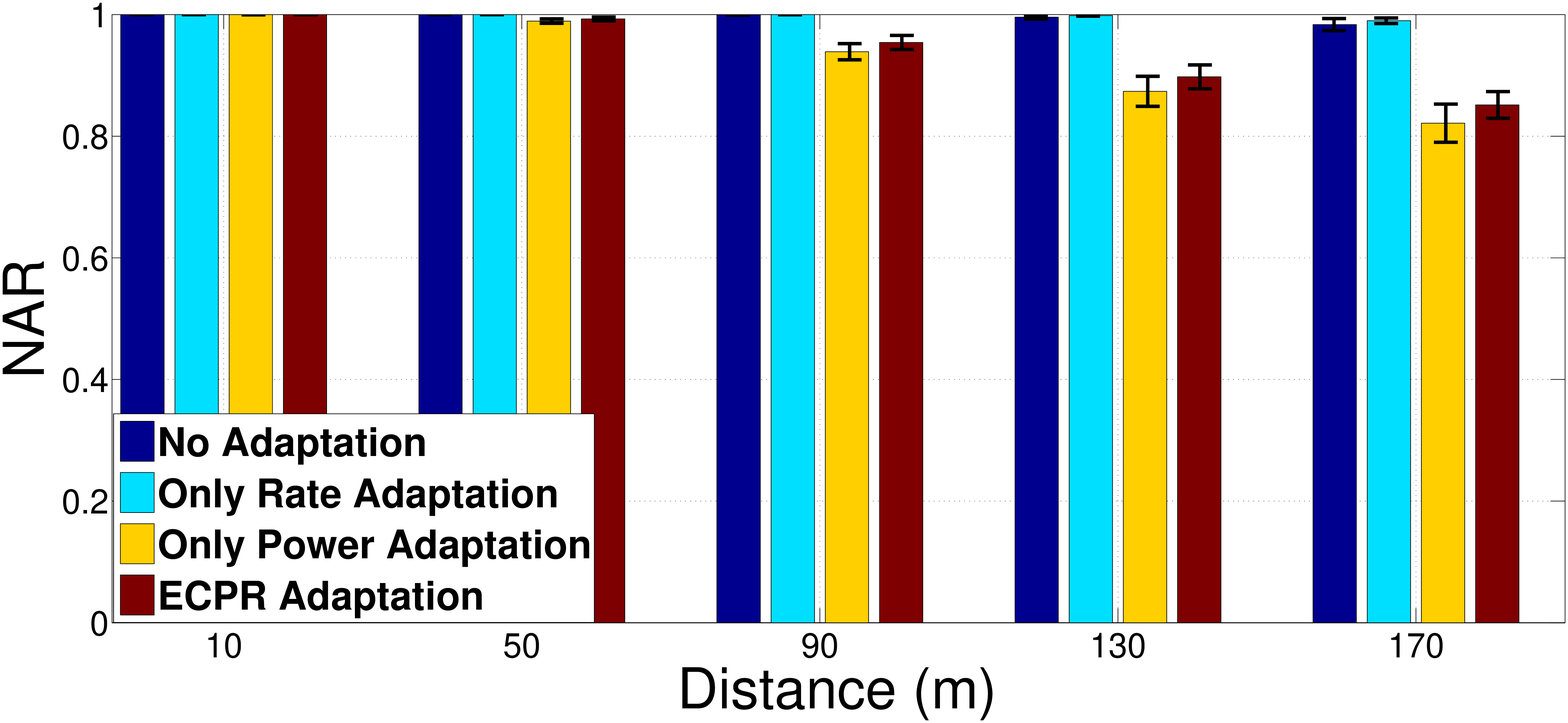}}\hspace{1mm}
\subfigure[RNAR]{\label{rnarUrban50TxPwr23}\includegraphics[width=0.49\textwidth]{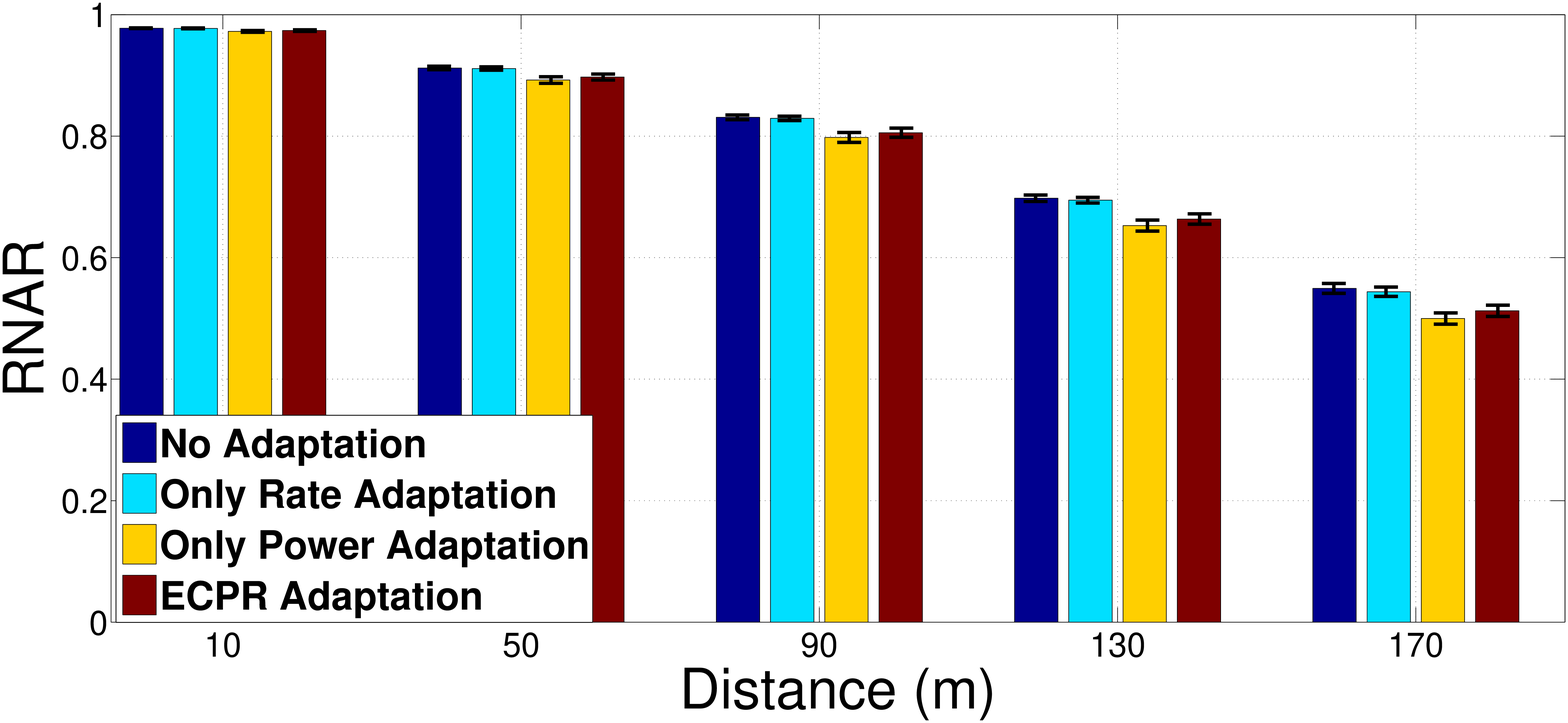}}\hspace{1mm}  
  \subfigure[Message rate]{\label{rateUrban50TxPwr23}\includegraphics[width=0.49\textwidth]{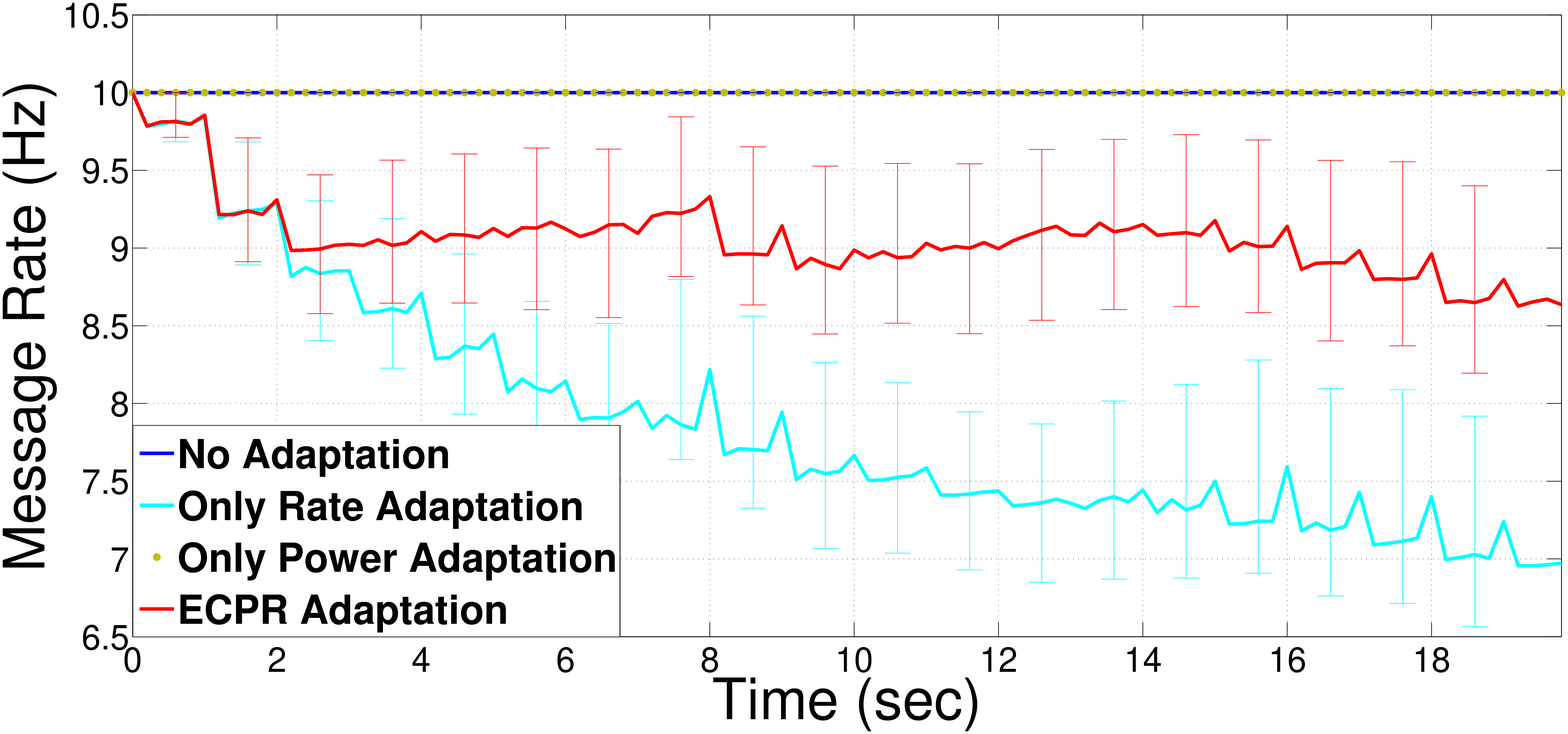}}\hspace{1mm}
 \subfigure[Transmit power]{\label{powerUrban50TxPwr23}\includegraphics[width=0.49\textwidth]{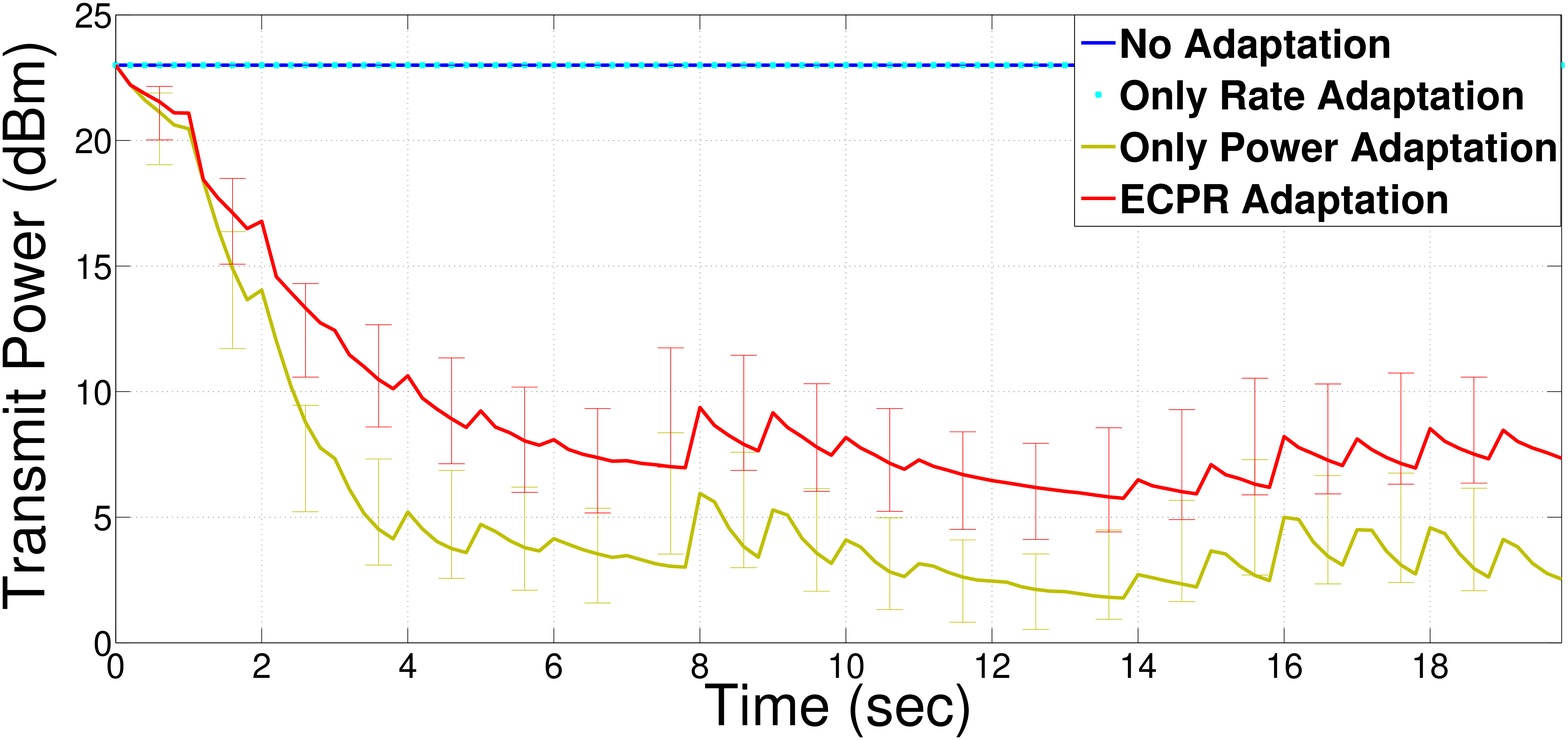}}\hspace{1mm}
 \subfigure[CBR]{\label{cbrUrban50TxPwr23}\includegraphics[width=0.49\textwidth]{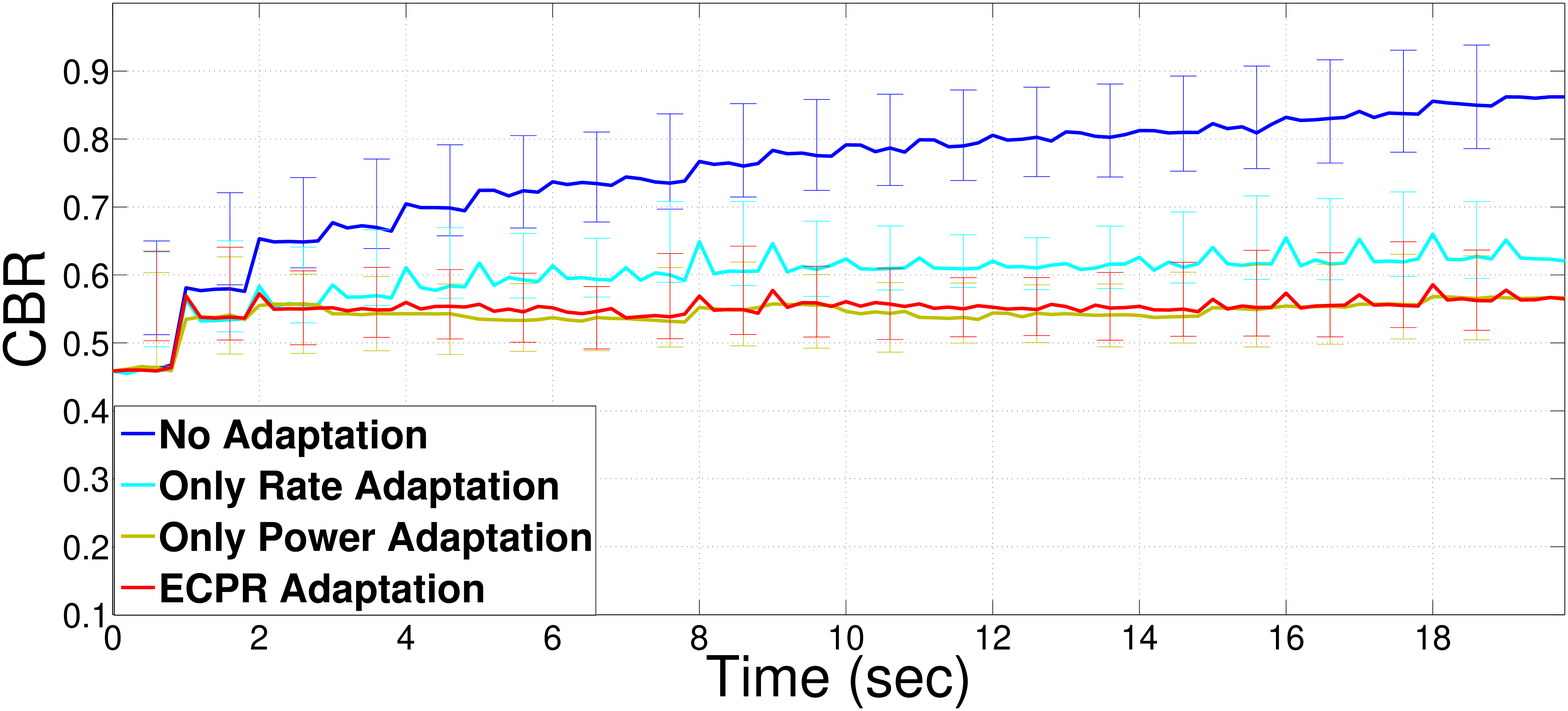}}\hspace{1mm}
     \caption{Target Awareness $85\%$, Target Awareness Distance = 50m, default Tx Power = 23 dBm. Urban Scenario. In this application context, ECPR can reduce the average power while not jeopardizing awareness. This allows for increase of overall throughput in the system as visible through increased average rate, while at the same time keeping the average CBR lower than that of rate-only algorithm. 
     }  
      \label{powerRateDiffAdp50_2}
   \end{center}
\end{figure*}

Figure~\ref{powerRateDiffAdp150} shows the results for the urban environment with a target awareness range of $150~m$, a default transmit power of $10~dBm$. Compared to rate-only (LIMERIC), ECPR can achieve a $20\%$ increase in points better awareness at the target distance by reducing the average rate from approximately $9~Hz$ to $8~Hz$. This scenario can be regarded as awareness-focused, where an application (\textit{e.g.}, intersection collision detection) requires vehicles to be aware of other vehicles within $150~m$ range. In this case, it is reasonable to trade some of the rate to increase the transmit power (Figure~\ref{powerUrban150TxPwr10}) and obtain an overall better awareness, since the messages that are traded for increased awareness are likely cooperative awareness messages at lower power, which would not be able to reach all vehicles at desired range, which defeats the purpose of sending those messages in the first place. Power-only algorithm achieves awareness (NAR) comparable to ECPR; however, due to not taking channel load (CBR) into account, it would exceed the target CBR. 

Figure~\ref{powerRateDiffAdp50_2} shows results for an urban environment with target awareness range of $50~m$, default transmit power of $23~dBm$ and showing how ECPR can achieve up to $25\%$ better average message rate, \bengi{for} the same satisfying requirement of the awareness rate at target awareness range. In this scenario, because the application context allows it, ECPR can reduce the average power (Figure~\ref{powerUrban50TxPwr23}) while not jeopardizing awareness. This allows for an increase of overall throughput in the system (see Figure~\ref{rateUrban50TxPwr23}), while at the same time keeping the average CBR lower than that of rate-only algorithm (see Figure~\ref{cbrUrban50TxPwr23}). In this scenario, no DCC adaptation performs as well as rate-only in terms of awareness; however, the CBR target is not satisfied. This emphasizes the need for DCC algorithms, since without adaptation there is a risk of channel overload and communication breakdown in case of high vehicular density. Note that ECPR can only adapt to awareness and rate requirements to the extent allowed by the physical surroundings (\textit{e.g.}, it is not possible to reach $500~m$ awareness range with $95\%$ awareness rate without very high transmit power) and transmit power parameters (which we limit to $0-23~dBm$ range so as to comply with the capabilities of existing IEEE 802.11p radios). 

\begin{figure*}[t!]
  \begin{center}
 \subfigure[Channel busy ratio]{\label{urbanCbrRandom100}\includegraphics[width = 0.49\textwidth]{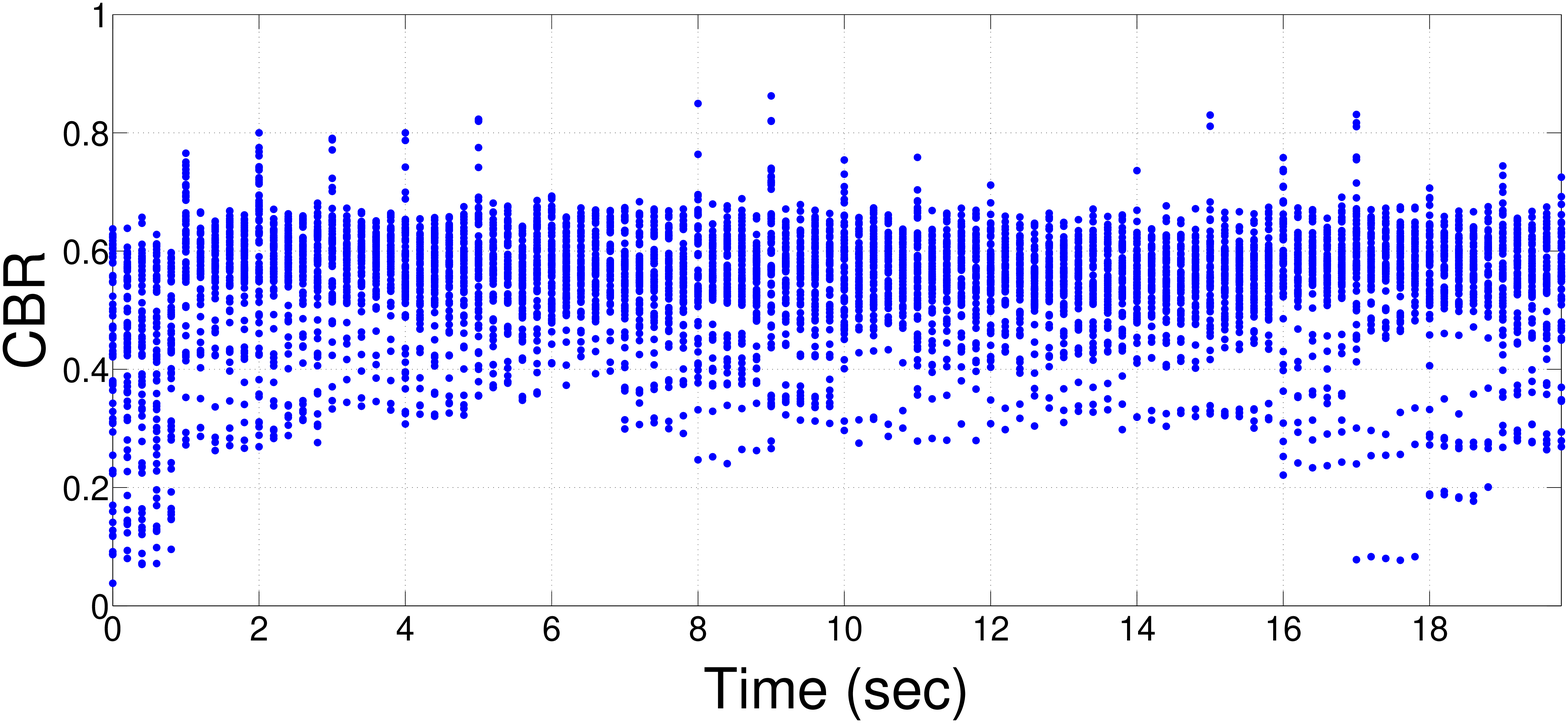}}\hspace{1mm}
 \subfigure[Message rate]{\label{urbanRateRandom100}\includegraphics[width=0.49\textwidth]{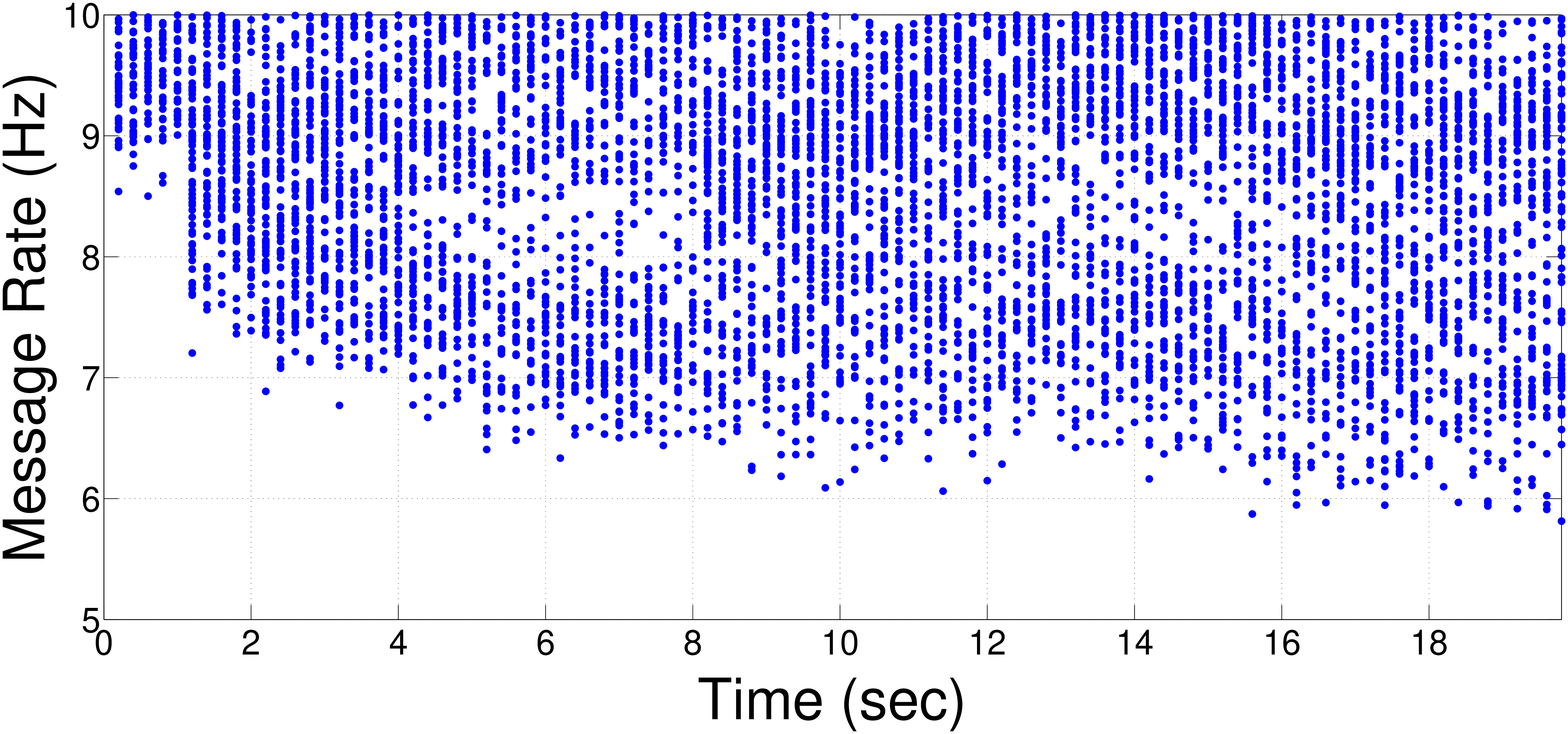}}\hspace{1mm}
\caption{Randomly selected 100 vehicles for Target Awareness Distance = 50~m, default Tx Power = 23~dBm.}
\label{urbanRandom100scattering}
   \end{center}
\end{figure*}

In Figure~\ref{urbanRandom100scattering} the per-vehicle behavior of the CBR and rate for 100 randomly chosen vehicles is shown. Although CBR overshoots the threshold CBR at each time step for both scenarios, it happens for one time step only, specifically when new vehicles enter the simulation. In the next step, the ECPR adapts the beacon rates to keep the CBR under the threshold. Regarding per-vehicle statistics, the results show that ECPR can control the load and can meet the target rate for all vehicles whose awareness requirements and environment allow it. It is important to note that ECPR aims to reach both the target awareness range and message rate based on the application requirements and given the constraints of specific physical environment. This results in a relatively large message rate spread, since the environment dictates that some vehicles need to transmit at higher power to reach the neighbors to which it has a bad channel (e.g., those behind a corner), which in turn increases the load for those neighbors to which it has a good (LOS) channel. In other words, combined awareness and rate control will not result in the same message rate at all vehicles unless their propagation environment is the same. 
\begin{figure*}[t!]
  \begin{center}
 \subfigure[The number of vehicles that can achieve the target message rate for Target Awareness Distance = 50~m, default Tx Power = 23~dBm.]{\label{reachedTarRate50TxPwr23}\includegraphics[width=0.49\textwidth]{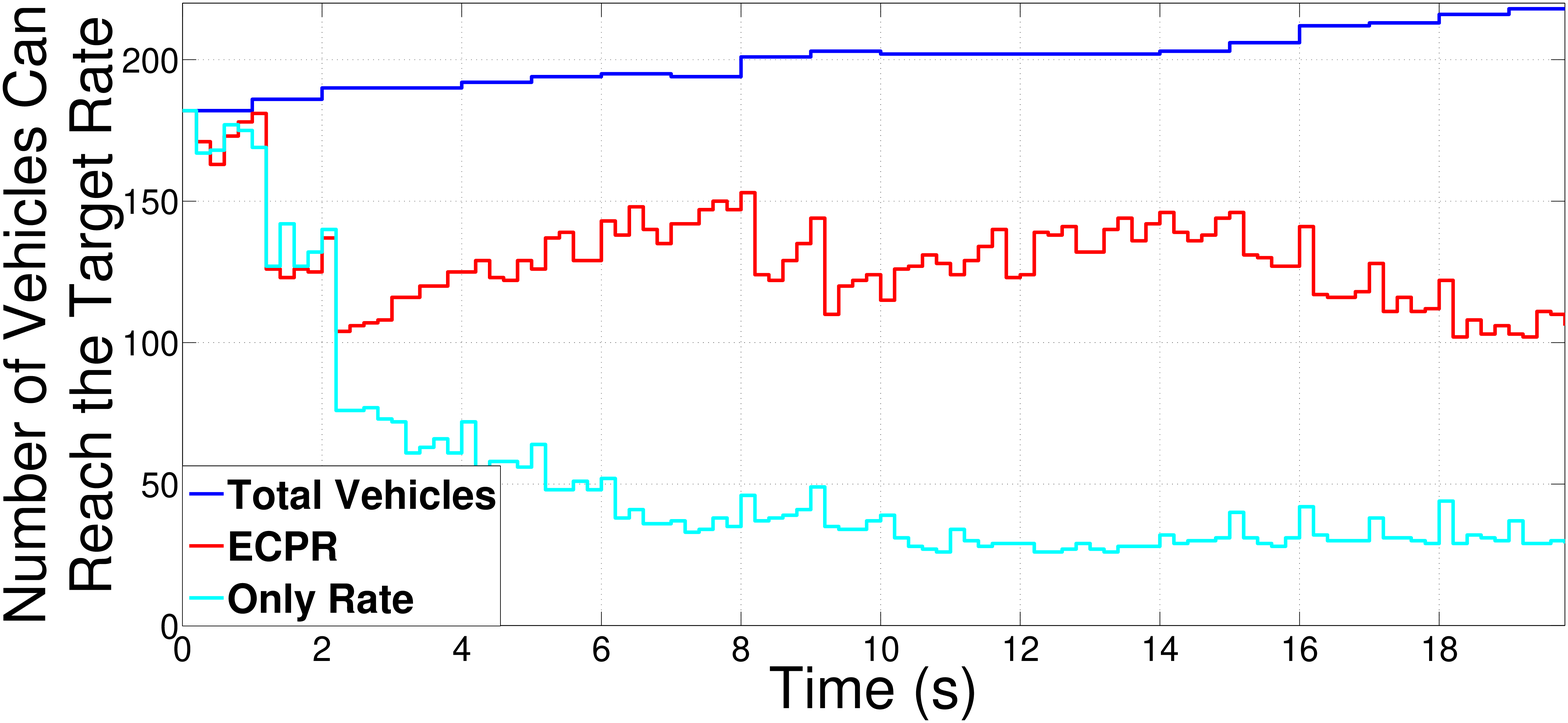}}\hspace{1mm}
   \subfigure[The number of vehicles that can achieve the target awareness. Target Awareness Distance = 150~m; default Tx Power = 10~dBm.]{\label{reachedTarAwar150TxPwr10}\includegraphics[width=0.49\textwidth]{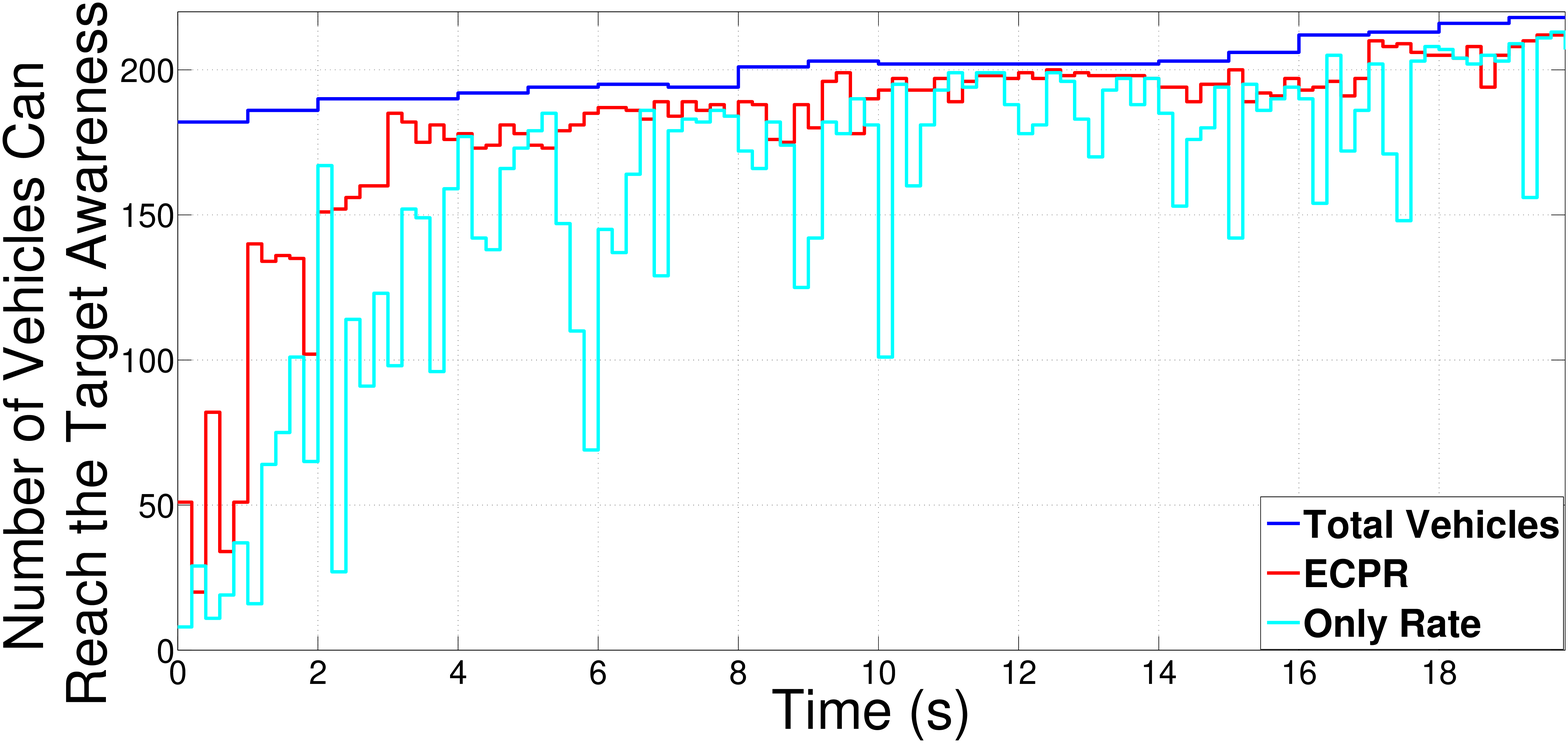}}\hspace{1mm}
  \caption{The number of vehicles that can achieve the target awareness.The number of vehicles that can reach awareness target, $85\%$, and rate target, $10~Hz$, for rate-only algorithm and ECPR. As a result of adaptation on transmission power on ECPR,  frequency reuse is able to be used more actively,  more vehicles reach the target message rate, and reaches target awareness more stably than rate-only adaptation.}
   \end{center}
   \label{vehiclesCanReachTarget}
\end{figure*} 

In Figure~\ref{reachedTarRate50TxPwr23} the number of vehicles that can achieve the target message rate, $10~Hz$ for this experiment, is shown for rate-only and ECPR adaptations. Since ECPR adapts the transmission power to various context,  transmission power is reduced if needed.  As a result of adaptation on transmission power,  frequency reuse is able to be used more actively and more vehicles reach the target message rate than rate-only adaptation. In addition to target rate, the number of vehicles that can achieve the awareness target, $85\%$, is compared in Figure~\ref{reachedTarAwar150TxPwr10}. Rate-only adaptation uses default transmission power therefore has limited capability to achieve target awareness for any kind of application while ECPR can adapt the transmission power to changing application and environment. Consequently, ECPR reaches target awareness more stably than rate-only adaptation.

\begin{figure*}[t!]
  \begin{center}
     \subfigure[Highway scenario average message  rate]{\label{rateHighwayTest}\includegraphics[width=0.49\textwidth]{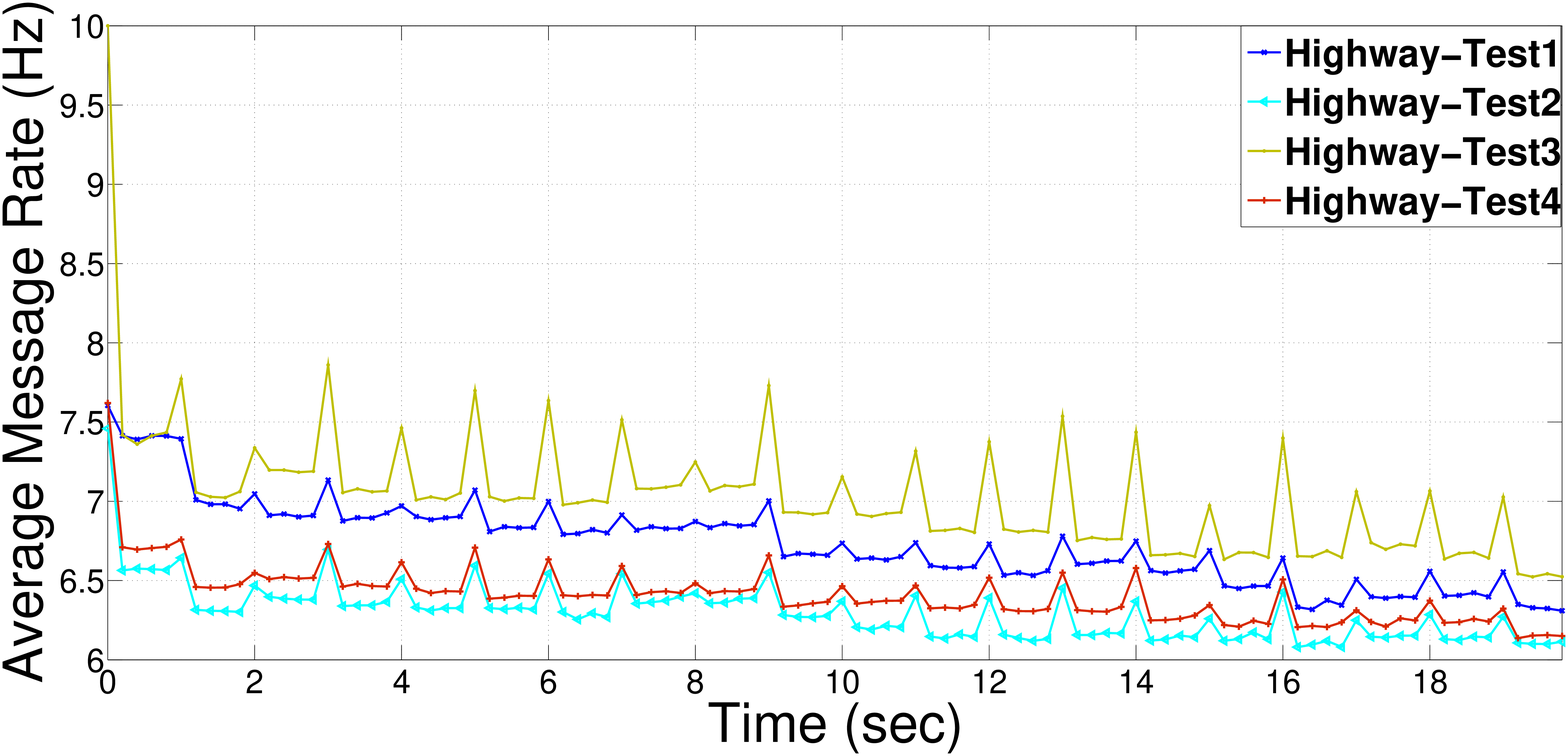}}\hspace{1mm}
  \subfigure[Highway scenario average transmit power]{\label{powerHighwayTest}\includegraphics[width=0.49\textwidth]{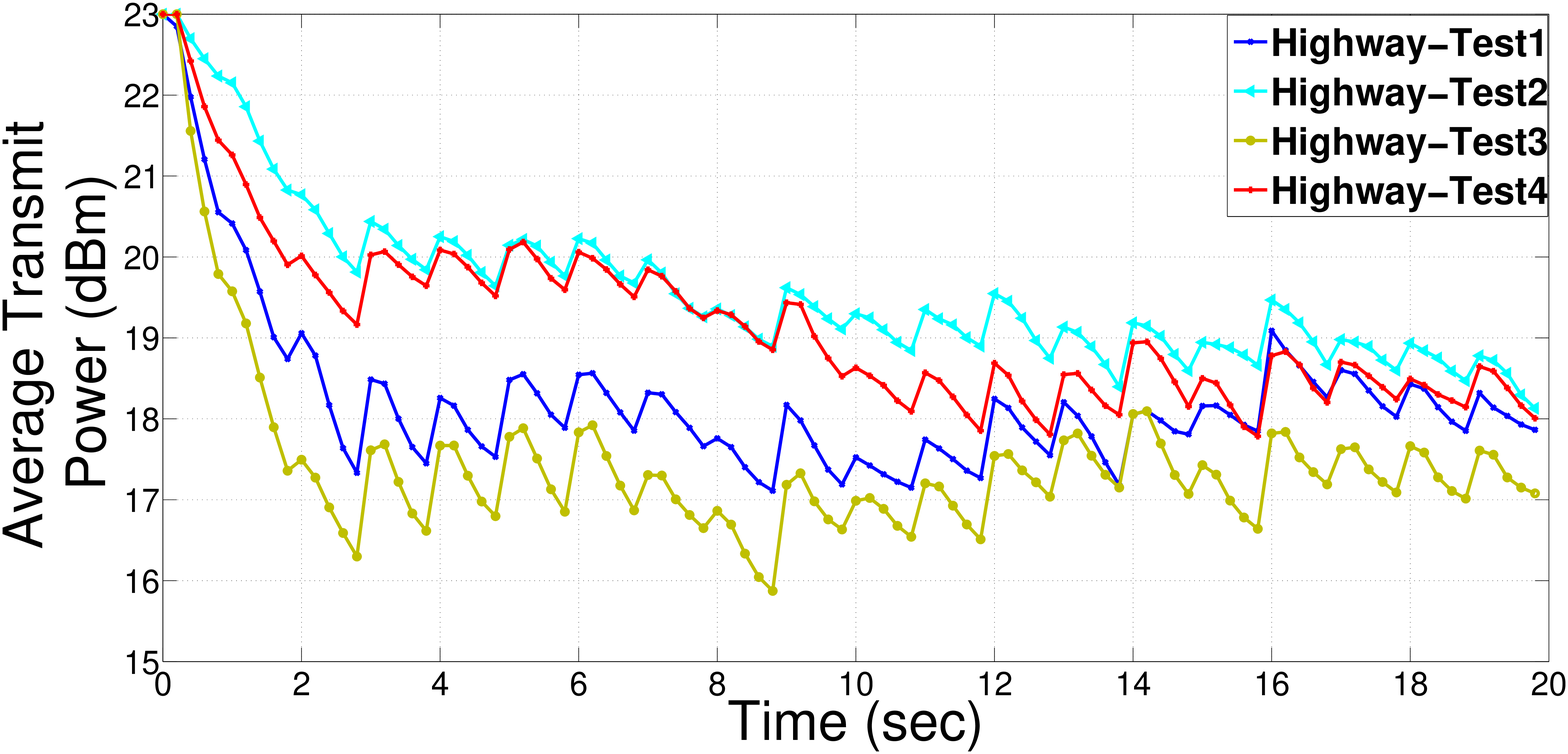}}\hspace{1mm}
 \subfigure[Urban scenario average message  rate]{\label{rateUrbanTest}\includegraphics[width=0.49\textwidth]{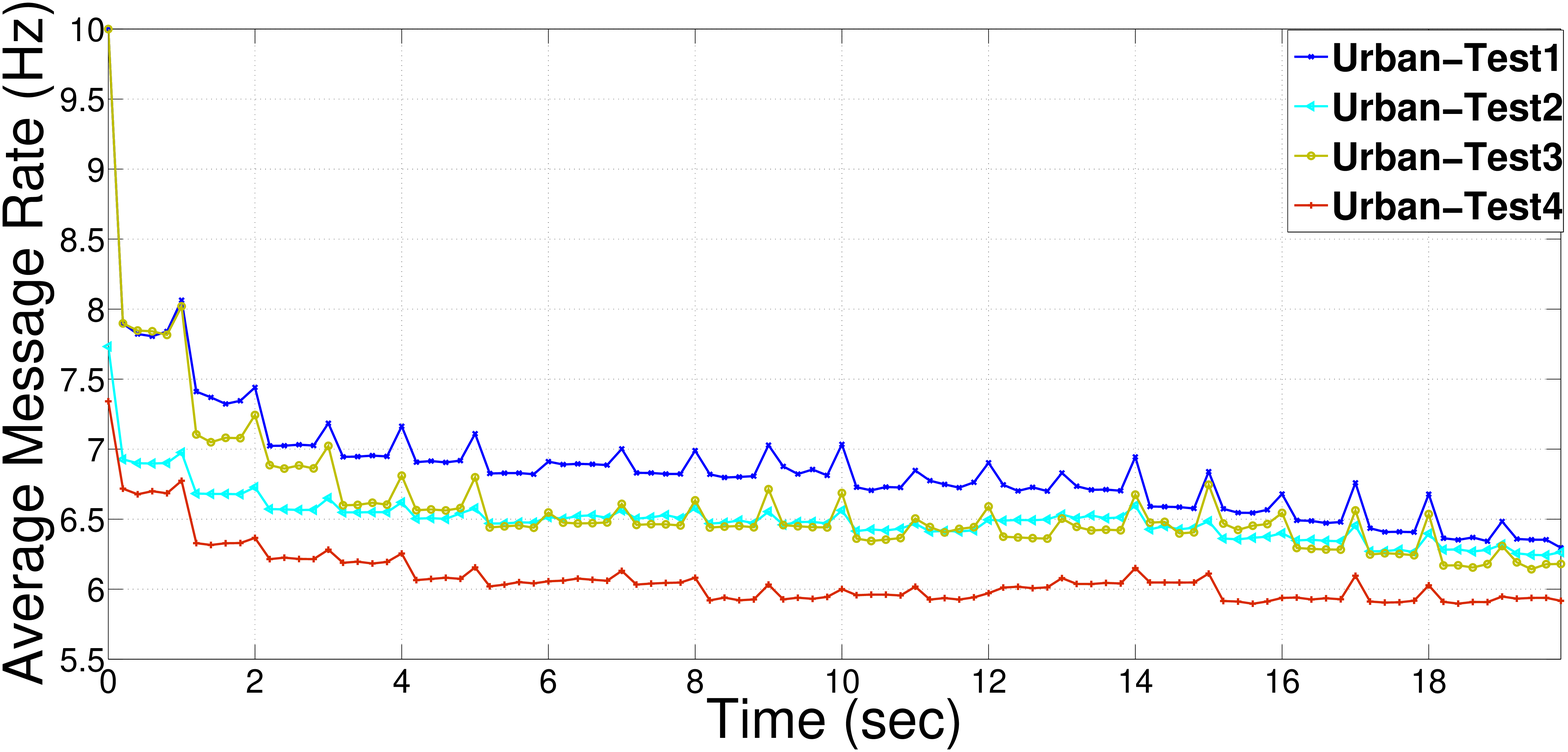}}\hspace{1mm}
    \subfigure[Urban scenario average transmit power]{\label{powerUrbanTest}\includegraphics[width=0.49\textwidth]{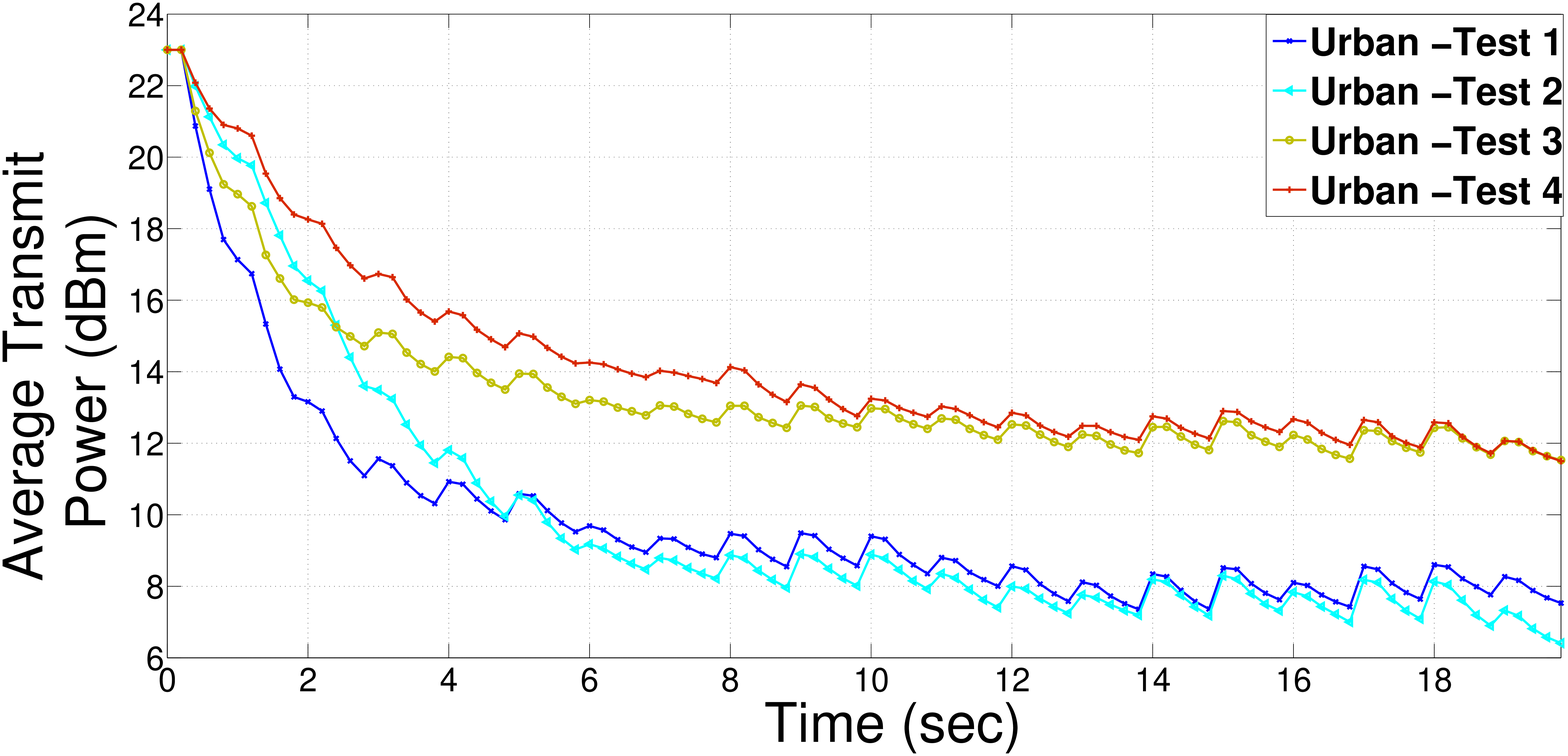}}\hspace{1mm}
     \caption{Average transmit Power and beacon rate for highway and urban environments. The relationship between average message rate and average transmit power is reversely proportional on each environment. }  
      \label{powerRateTest}
   \end{center}
\end{figure*}

ECPR is tested for different default transmission power values to see its adaptation ability to any environment and context cases.  However,  we use 10 dBm power and 150 m target range (low default power, high range requirement) and 23 dBm power and 50 m target range (high default power, low range requirement) to show how ECPR performs in comparatively extreme cases.

\subsection{Different Target Rate and Awareness Distance Sets for Combined Algorithm: Urban vs Highway Environment}

Figure~\ref{powerRateTest} shows average \bengi{message rates and transmit powers} for different tests. Target awareness range and message rate are denoted in Table~\ref{tests}. The relationship between average message rate and average transmit power is reversely proportional on each environment: the lower the average power, the smaller the message coverage, resulting in better channel reuse and higher rate. The average rate is similar in the two environments because the high density of vehicles means that the channel is loaded most of the time. Interesting to note is that in urban scenarios, the average power converges to a value lower than in highway scenarios; this can be attributed to the increased number of neighbors for the same range in urban environment. Thus, the channel becomes more congested from neighbors at shorter distance and requiring lower power to reach them. In turn, this offsets the range limitations due to obstructing buildings requiring larger power for the same range at highways. 

\begin{figure*}[t!]
  \begin{center}
  \subfigure[Highway scenario]{\label{deltaRateHighwayTest}\includegraphics[width=0.49\textwidth]{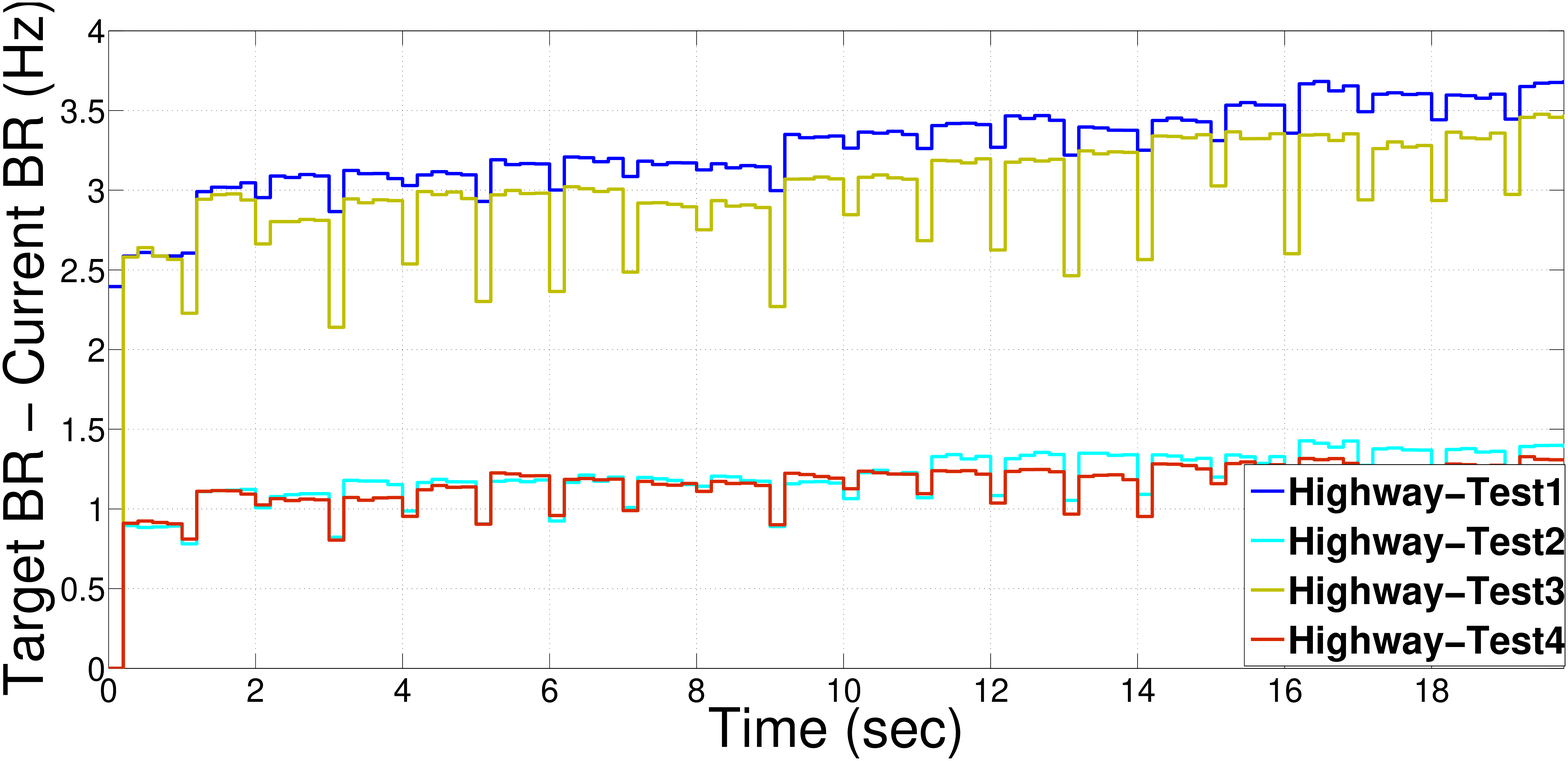}}\hspace{1mm}
   \subfigure[Urban scenario]{\label{deltaRateUrbanTest}\includegraphics[width=0.49\textwidth]{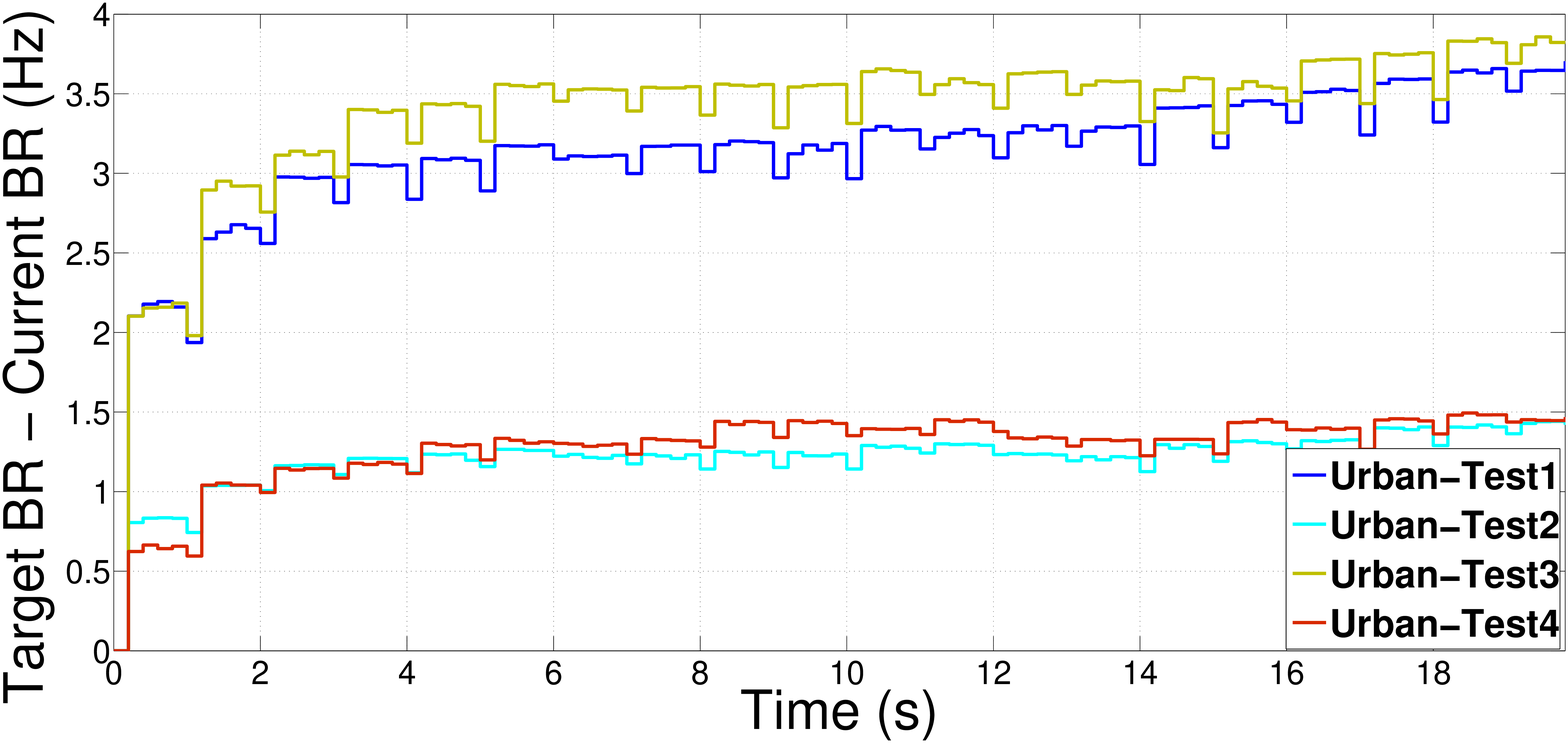}}\hspace{1mm}
     \caption{Average difference between target and achieved message rate for highway and urban environments. Test 1 and 3 target the maximum message rate, the difference between target and current rate is higher than in Test 2 and Test 4. The target rate is on average less than maximum  rate, thus the difference of achieved to target rate is less.}  
      \label{deltaRateTest}
   \end{center}
\end{figure*}
\begin{figure*}[t!]
  \begin{center}
  \subfigure[Highway scenario]{\label{CBRHighwayTest}\includegraphics[width=0.49\textwidth]{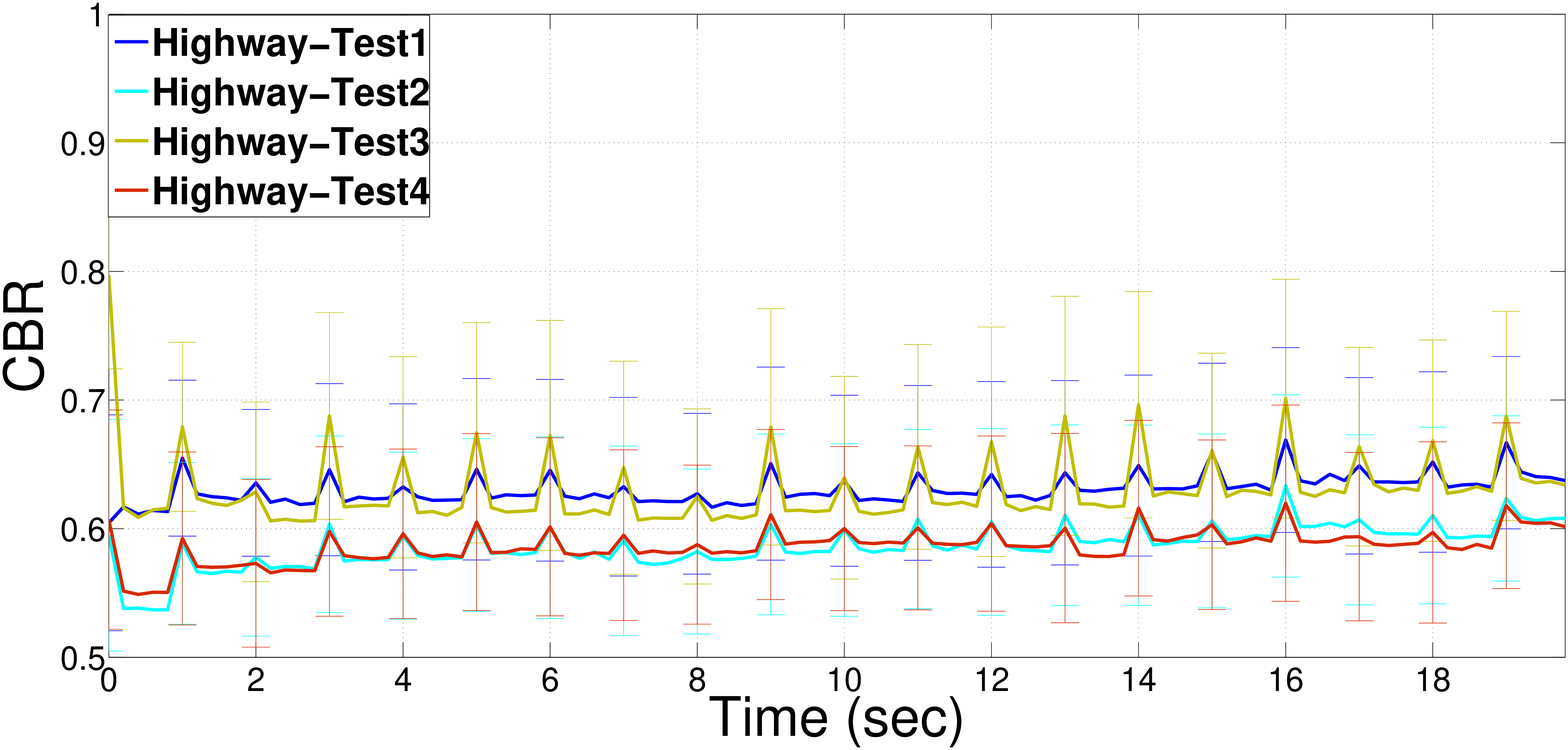}}\hspace{1mm}
    \subfigure[Urban scenario]{\label{CBRUrbanTest}\includegraphics[width=0.49\textwidth]{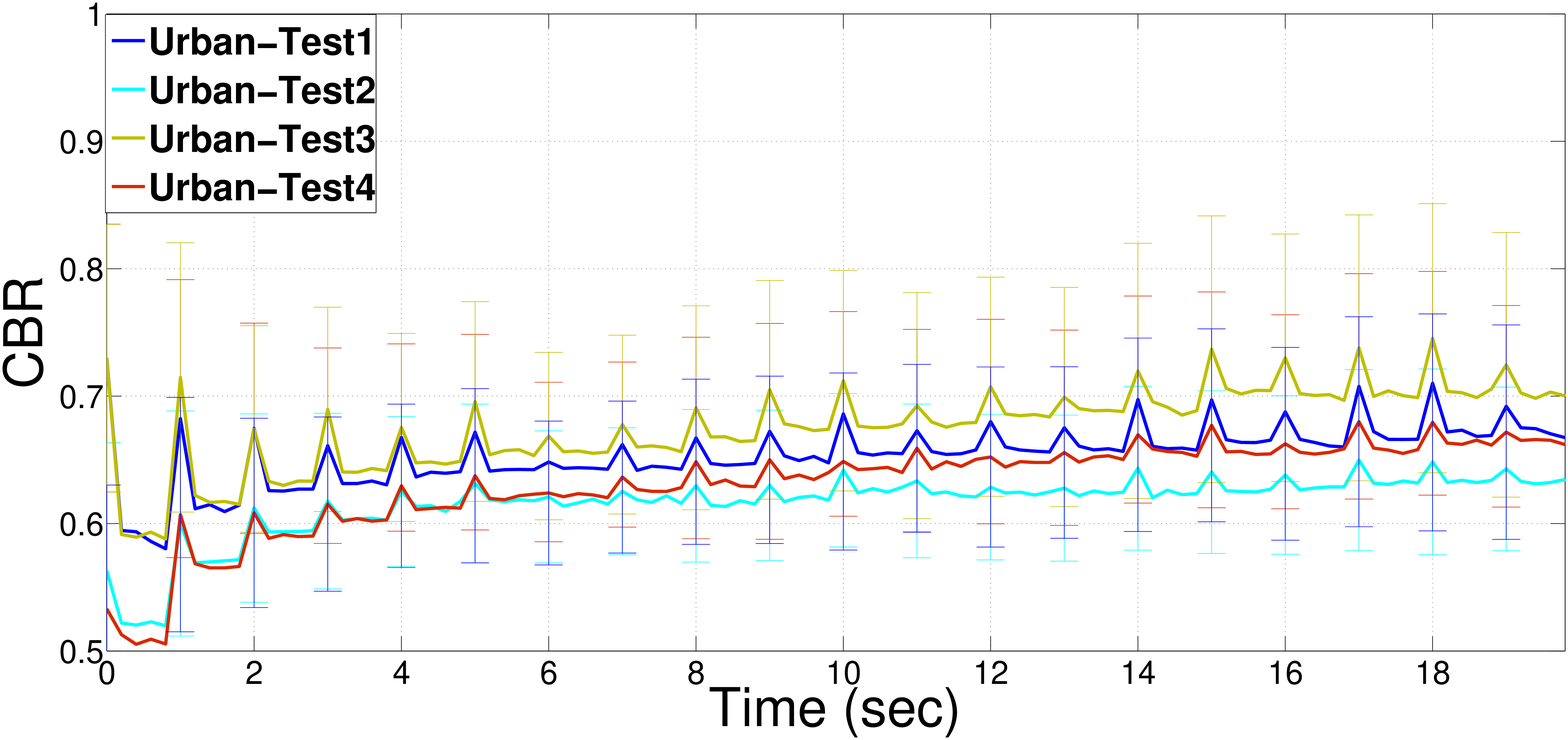}}\hspace{1mm}
     \caption{Standard deviation and mean of CBRs in highway and urban environments.   The threshold CBR value is set as 0.6 with $\pm$ 0.05 tolerance.In urban scenario, average CBR is higher than in the highway scenario. The reason is that each ego node needs to communicate with a larger number of neighboring vehicles in urban environment than highway due to the vehicles being concentrated around intersections~\cite{tonguz09_2}; combined with higher power to achieve the same awareness, this results in higher overall CBR.}  
      \label{CBRtests}
   \end{center}
\end{figure*}

Figure~\ref{deltaRateTest} shows the difference between the target message rates and the achieved rate for both urban and highway scenarios. Since Test 1 and 3 target the maximum message rate, the difference between target and current rate is higher than in Test 2 and Test 4. In other words, in Tests 2 and 4, the target rate is on average less than maximum  rate, thus the difference of achieved to target rate is less.

Figure~\ref{CBRtests} shows the average CBR levels and their standard deviations for each time step for all tests. As expected, the test which has higher average message rate also  has higher CBR values. However, average CBR values never overflow the CBR threshold, which is $0.6$ with $\pm0.05$ tolerance. Although new vehicles entering the simulation and starting at maximum transmit power join the communication at each second, ECPR adapts the power and message rate at the next time step and decreases the CBR to threshold value. In urban scenario, average CBR is higher than in the highway scenario. The reason is that each ego node needs to communicate with a larger number of neighboring vehicles in urban environment than highway due to the vehicles being concentrated around intersections~\cite{tonguz09_2}; combined with higher power to achieve the same awareness, this results in higher overall CBR. 

The results show that ECPR can effectively adapt the power and rate to achieve the target requirements on awareness and rate given by the application context, irrespective of the propagation environment. Since it has the ability to obtain higher average rate when the awareness requirements allow it, at the same time maintaining or reducing the CBR as compared to rate-only solution, it can be used to improve the overall system throughput. Conversely, if the awareness requirements are more stringent or the propagation environment more harsh, ECPR efficiently trades rate to improve the awareness.

\subsection{Effect of Medium Access Layer Collisions}
\begin{figure*}[t!]
  \begin{center}
     \subfigure[NAR]{\label{narMAC}\includegraphics[width=0.49\textwidth]{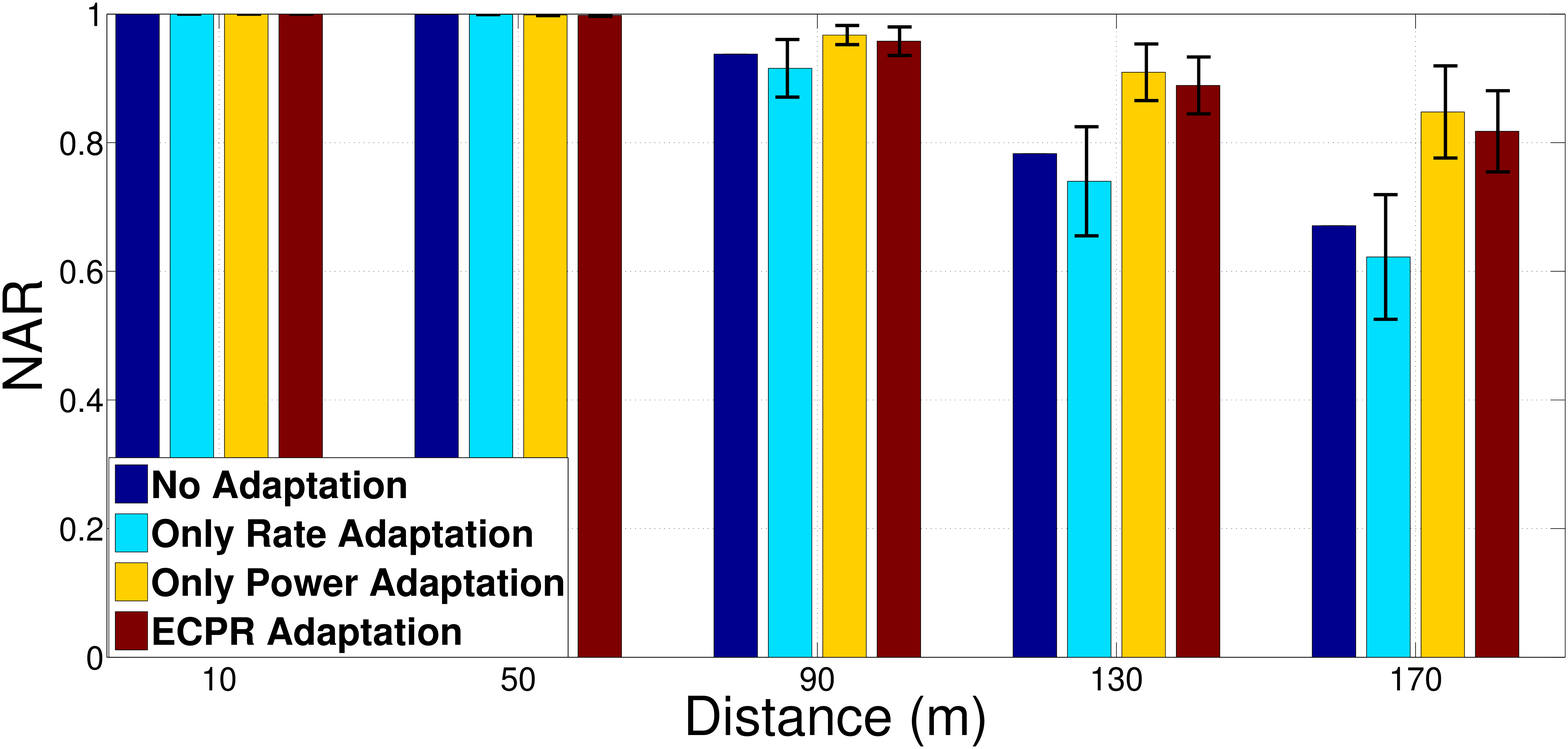}}\hspace{1mm}
\subfigure[RNAR]{\label{rnarMAC}\includegraphics[width=0.49\textwidth]{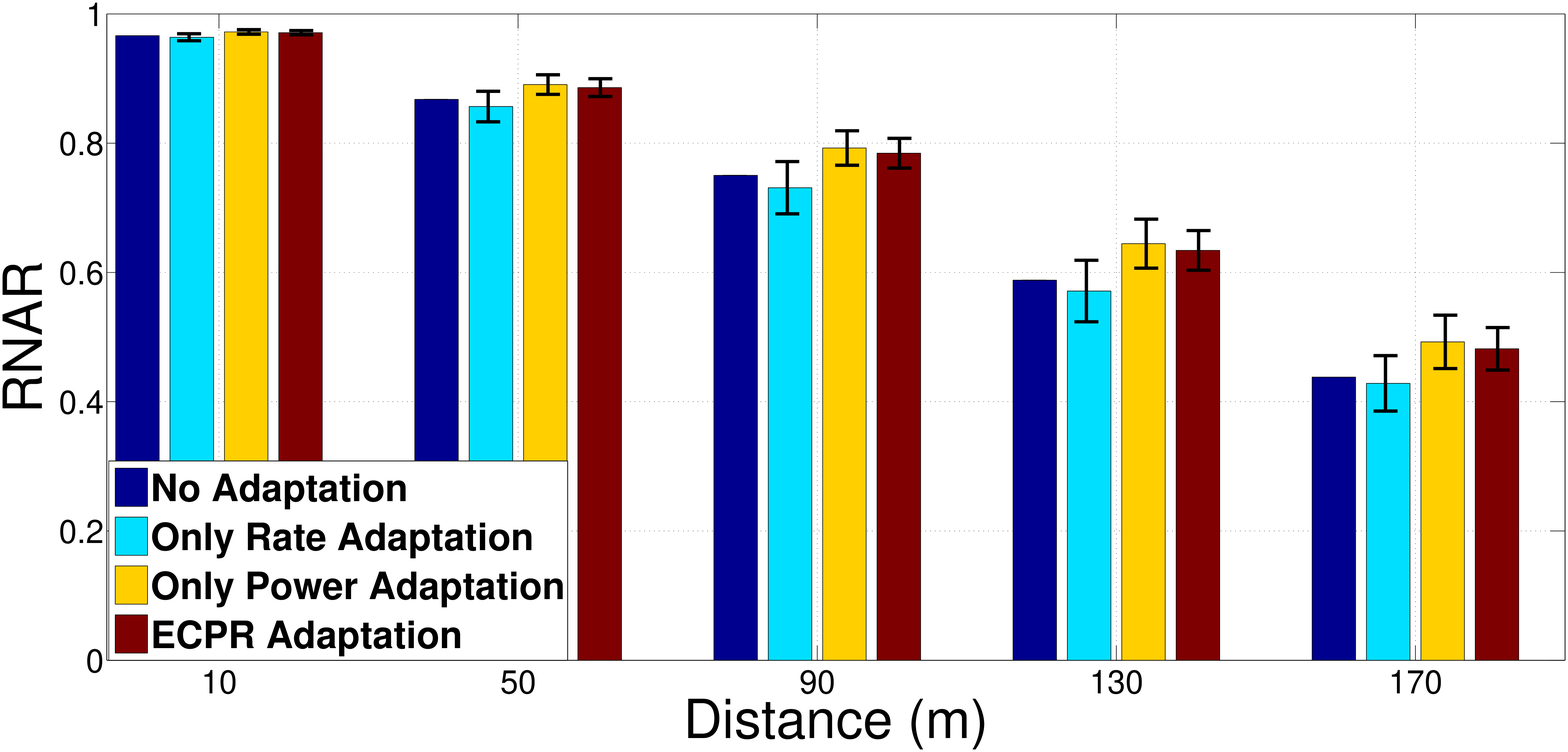}}\hspace{1mm}  
  \subfigure[Message rate]{\label{rateMAC}\includegraphics[width=0.49\textwidth]{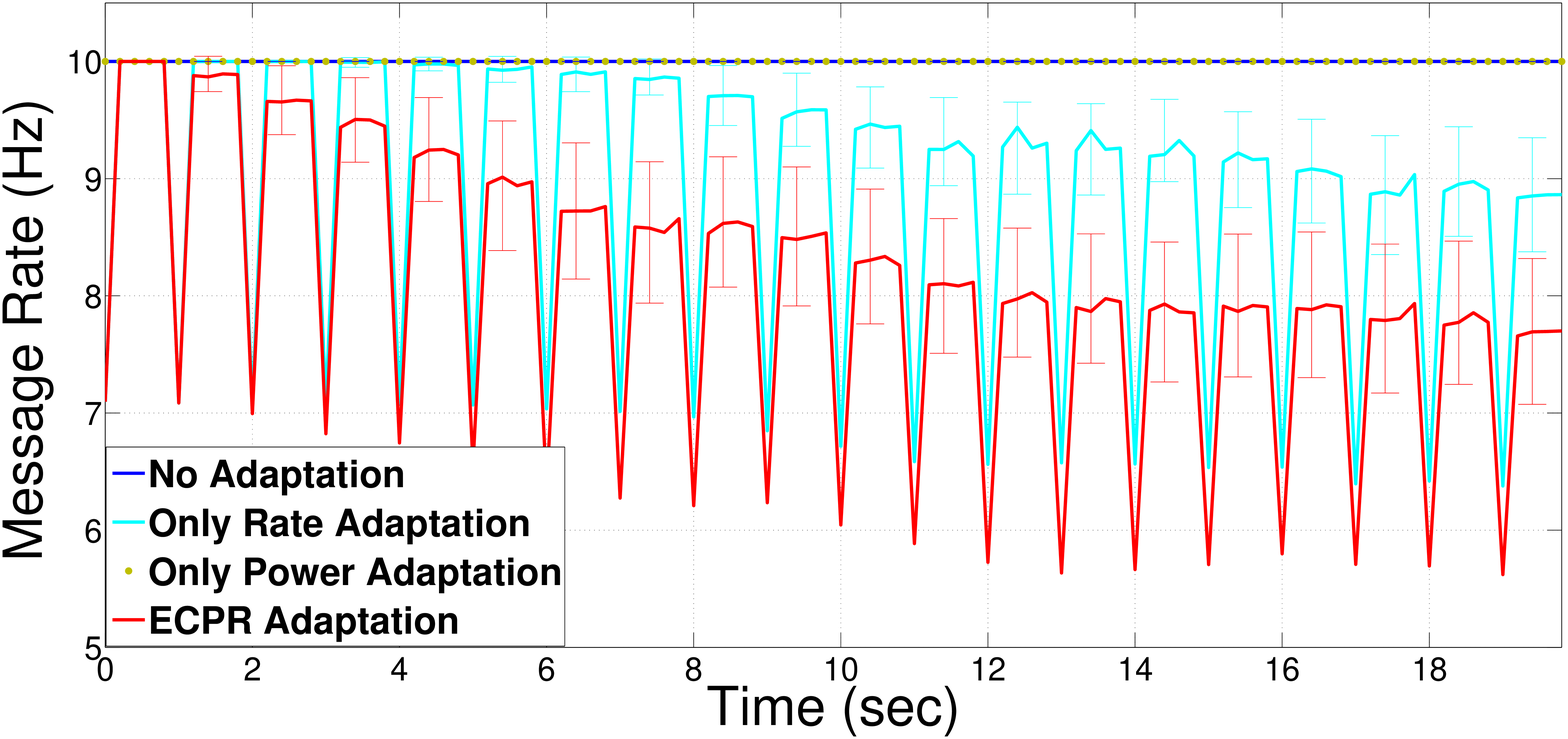}}\hspace{1mm}
 \subfigure[Transmit power]{\label{powerMAC}\includegraphics[width=0.49\textwidth]{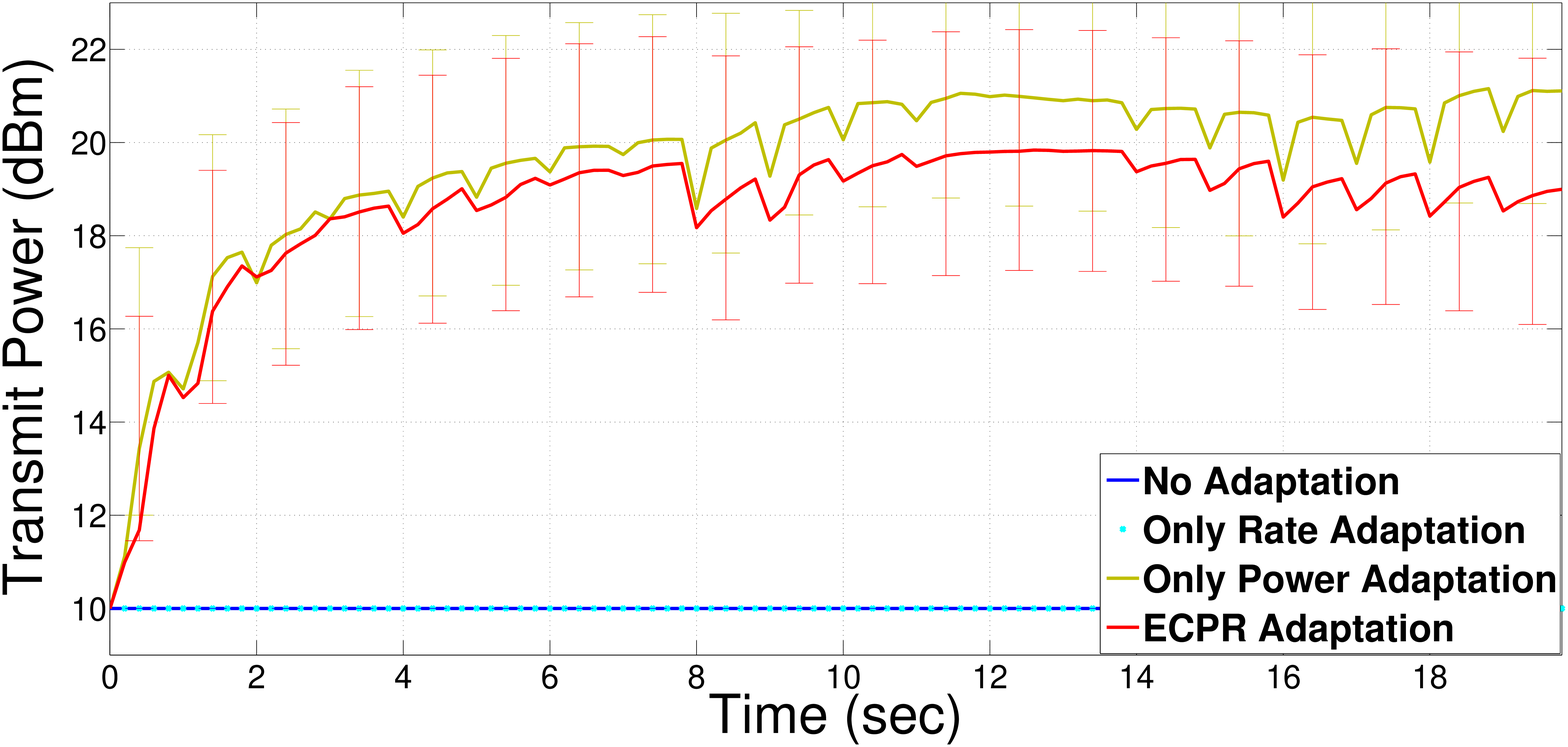}}\hspace{1mm}
 \subfigure[CBR]{\label{cbrMAC}\includegraphics[width=0.49\textwidth]{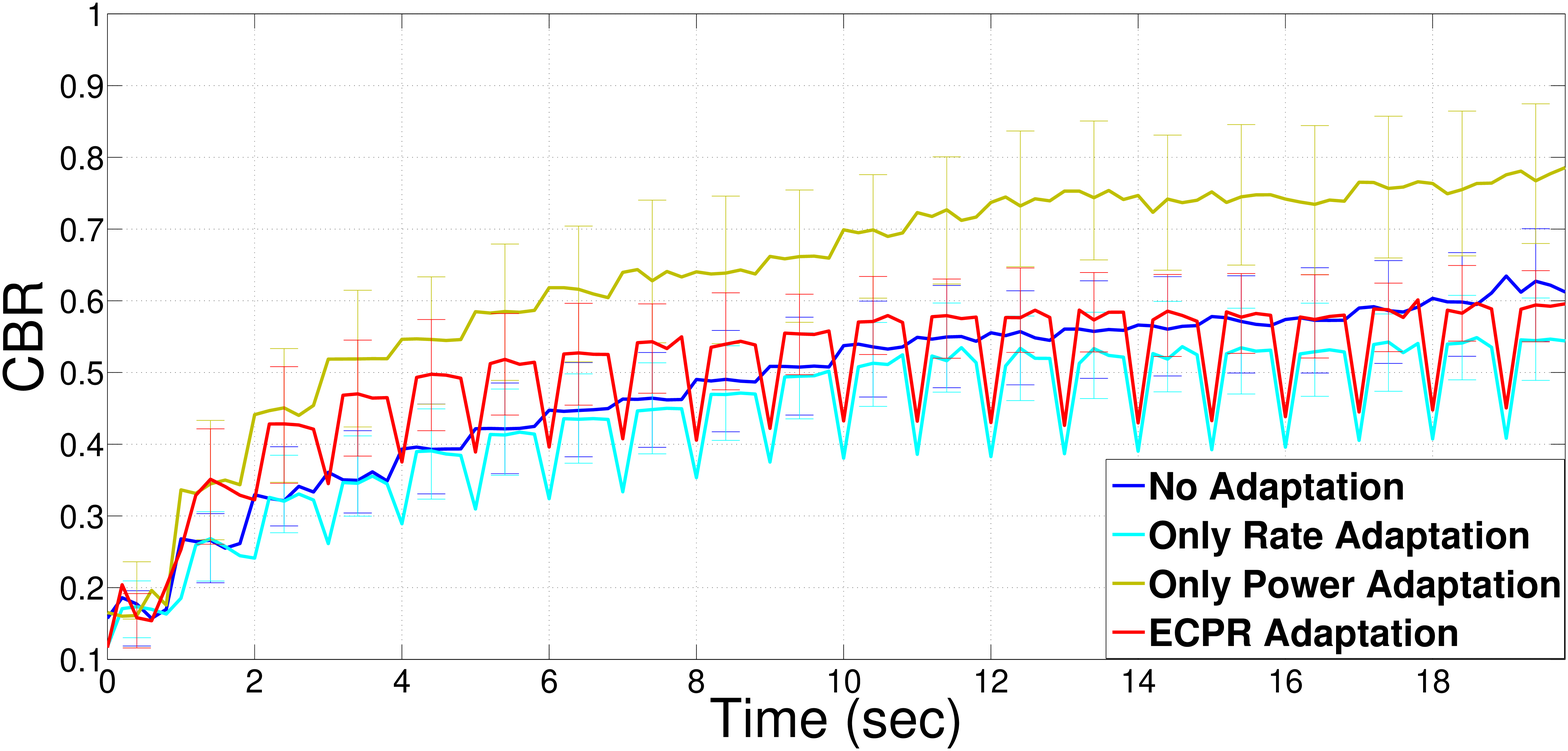}}\hspace{1mm}
     \caption{Target Awareness $85\%$, Target Awareness Distance = 50m, default Tx Power = 23 dBm. Urban Scenario with MAC collisions.}  
      \label{mac}
   \end{center}
\end{figure*}

 To investigate the effect of Medium Access Layer (MAC) collisions on the performance of ECPR, we perform simulations with the same network conditions as for the scenario shown in Figure~\ref{powerRateDiffAdp150} (Target Awareness 85\%, Target Awareness Distance = 150m, default Tx Power = 10 dBm), with increased loss due to MAC collisions (note that results in Figure~\ref{powerRateDiffAdp150} consider no loss due to MAC collision). The collision statistics are defined as follows: when CBR is below $20\%$, $20-30\%$, $30-40\%$, $40-50\%$, $50-60\%$, and above $60\%$, MAC layer collision causes  $0\%$, $1\%$,$3\%$,$7\%$, $10\%$, $30\%$ packets drops, respectively. These parameters are selected to represent harsh conditions caused by progressively increasing collisions with the increase in channel load~\cite{choi2003ieee}.
 Compared with Figure~\ref{powerRateDiffAdp150},  Figure~\ref{narMAC}--~\ref{rnarMAC}, shows that the effect of MAC collisions is quite limited in terms of the key performance metrics of ECPR (NAR, RNAR); similarly limited difference can be observed in Figures~\ref{rateMAC}--~\ref{cbrMAC} in terms of the resulting network parameters (message rate, transmit power, and CBR). Therefore, we conclude that ECPR utilizes channel as effective as possible while keeping CBR under the threshold even in the face of MACe collisions. \bengi{In~Figure~\ref{rateMAC}, the dip points are how network parameters react to changes without any adaptation yet. The ECPR adapts the parameters to the optimum values  every $200~msec$ by considering the resource limitations.} 
  
\begin{table}[!t]
\centering
\small
\caption{Average percentage of potentially hidden nodes for ECPR and rate-only (LIMERIC) algorithm.} 
\vspace{-5pt}
\begin{center}
\begin{tabular}{|l |c |c |c |c|}
\hline
  \multirow{3}{*}{~}				& \multicolumn{2}{|c|}{Transmit Power = 23~dBm}    &  \multicolumn{2}{|c|}{Transmit Power = 10~dBm} \\ 
 												& \multicolumn{2}{|c|}{ Awareness Range = 50~m}   &  \multicolumn{2}{|c|}{Awareness Range = 150~m} \\ 
												& 		50 Vehicle		& 100 Vehicles       					   & 50 Vehicles					& 100 Vehicles \\ \hline
ECPR 										& 		12.9$\%$			&  	22.4$\%$						  &   8.5$\%$                    &		 17.4$\%$  \\
LIMERIC 									& 		11.9$\%$		&	    23.2$\%$       					  &  8.7$\%$                        &16.5$\%$	 \\	\hline								
\end{tabular}
\end{center}
\label{hnpTable}
\end{table}

Hidden node problem is another access layer consideration that can be caused by the propagation environment layout as well the transmit power variations. To illustrate the issue, consider the scenario in Figure~\ref{SysArchUrban}, where two vehicles on perpendicular roads are trying to transmit to vehicle in the center of intersection; if those two vehicles cannot ``hear'' each other,  they create the hidden node problem on the vehicle in the intersection. For each of A's neighbors, we check if that neighbor can ``hear'' from A's other neighbors. Each pair of A's neighbors that cannot hear each other is counted as potentially causing a hidden node problem at A. Thus, the percentage of hidden nodes is computed as the proportion of potentially hidden node pairs to total number of communication pairs. The results in Table~\ref{hnpTable} show that ECPR results in comparative percentage of hidden node pairs as LIMERIC (i.e., ECPR does not increase the probability of hidden nodes). 

\section{Conclusions}
\label{Conclusion}
In this paper, we proposed a combined rate and power DCC algorithm that efficiently achieves the target  awareness and rate requirements given by the application context (\textit{e.g.}, target applications, vehicle speed, traffic density) in varying propagation environments. By using path loss exponent estimation, ECPR adapts the transmit power to reach the target awareness range. ECPR controls the channel load by adjusting the rate and power according to the current channel load, awareness range, and rate information. We perform realistic simulations, incorporating real world information about mobile and static objects (vehicles, buildings, and foliage) and test ECPR in scenarios with varying LOS conditions, highly dynamic network topology, and different environments (highway and urban). We show that ECPR has the ability to obtain higher rate when the awareness requirements allow it, improving the average rate by 15+\%, while keeping the target awareness and channel load. If the awareness requirements are more stringent or the propagation environment more harsh, ECPR efficiently trades rate to improve the awareness by up to $20$ percentage points. ECPR can be implemented atop existing DCC solutions with little effort, as the only additional information it requires is the transmit power of the message that can be piggybacked in the message itself. 

\bibliographystyle{elsarticle-num} 
 \bibliography{referencesComcom}
\end{document}